\RequirePackage{fix-cm}

\documentclass[smallcondensed, draft, envcountsect, envcountreset]{svjour3}
\smartqed  
\usepackage{tkz-graph}
\usepackage{verbatim}
\usepackage{graphicx}
\usepackage{caption}

\spnewtheorem{Definition}{Definition}[section]{\bf}{\it}
\spnewtheorem{Remark}{Remark}[section]{\bf}{\it}
\spnewtheorem{Proposition}{Proposition}[section]{\bf}{\it}
\spnewtheorem{Property}{Property}[section]{\bf}{\it}
\spnewtheorem{Problem}{Problem}[section]{\bf}{\it}
\spnewtheorem{Theorem}{Theorem}[section]{\bf}{\it}
\spnewtheorem{Corollary}{Corollary}[section]{\bf}{\it}
\spnewtheorem{Lemma}{Lemma}[section]{\bf}{\it}

\def\done{\qed}

\newcommand{\bform}{\begin{displaymath}}
\newcommand{\eform}{\end{displaymath}}
\newcommand{\beqn}[1]{\begin{equation}\label{#1}}
\newcommand{\eeqn}{\end{equation}\smallskip}
\newcommand{\beqna}[1]{\begin{eqnarray}\label{#1}}
\newcommand{\eeqna}{\end{eqnarray}\smallskip}
\newcommand{\beqnan}{\begin{eqnarray*}}
\newcommand{\eeqnan}{\end{eqnarray*}\smallskip}

\usepackage[english]{babel}
\journalname{Mathematical Programming A}

\begin{document}

\title{An ${\cal O}(n^2 \log{n})$ algorithm for the weighted stable set problem in claw-free graphs}
\dedication{Dedicated to the memory of Manfred Padberg}
\titlerunning{Weighted stable set problem in claw-free graphs}

\author{Paolo Nobili \and Antonio Sassano}

\institute{P. Nobili \at
              Dipartimento di Ingegneria dell'Innovazione, Universit\`a del Salento, Lecce, Italy \\
           \and
           A. Sassano \at
              Dipartimento di Ingegneria Informatica, Automatica e Gestionale ``Antonio Ruberti'',  Universit\`a di Roma ``Sapienza'', Roma, Italy
}

\date{}

\maketitle


\begin{abstract}
A graph $G(V, E)$ is \emph{claw-free} if no vertex has three pairwise non-adjacent neighbours. The Maximum Weight Stable Set (MWSS) Problem in a claw-free graph is a natural generalization of the Matching Problem and has been shown to be polynomially solvable by Minty and Sbihi in 1980. In a remarkable paper, Faenza, Oriolo and Stauffer have shown that, in a two-step procedure, a claw-free graph can be first turned into a quasi-line graph by removing strips containing all the irregular nodes and then decomposed into \emph{\{claw, net\}-free} strips and strips with stability number at most three. Through this decomposition, the MWSS Problem can be solved in ${\cal O}(|V|(|V| \log |V| + |E|))$ time. In this paper, we describe a direct decomposition of a claw-free graph into \emph{\{claw, net\}-free} strips and strips with stability number at most three which can be performed in ${\cal O}(|V|^2)$ time. In two companion papers we showed that the MWSS Problem can be solved in ${\cal O}(|E| \log |V|)$ time in claw-free graphs with $\alpha(G) \le 3$ and in ${\cal O}(|V| \sqrt{|E|})$ time in \{claw, net\}-free graphs with $\alpha(G) \ge 4$. These results prove that the MWSS Problem in a claw-free graph can be solved in ${\cal O}(|V|^2 \log |V|)$ time, the same complexity of the best and long standing algorithm for the MWSS Problem in \emph{line graphs}.
\keywords{claw-free graphs \and quasi-line graphs \and stable set \and matching}
\end{abstract}

\section{Introduction}

\noindent
The \emph{Maximum Weight Stable Set (MWSS) problem} in a graph $G(V, E)$ with node-weight function $w: V \rightarrow \Re$ asks for a maximum weight subset of pairwise non-adjacent nodes. In a remarkable theoretical effort, Faenza, Oriolo and Stauffer \cite{FaenzaOS14} have proposed an elegant decomposition approach to the solution of the MWSS problem when $G$ is claw-free. The approach is based on a two-step decomposition technique and produces a family of \emph{strips}, a structure analogous to that introduced by Chudnowsky and Seymour \cite{ChudSeym} in their characterization of quasi-line graphs. In the first step the procedure in \cite{FaenzaOS14} removes a strip ``around'' each irregular node, as to end up with a quasi-line graph $\tilde G$. In the second step it performs the so called \emph{ungluing} operation to a special class of cliques of $\tilde G$, the \emph{articulation cliques}. The algorithm proceeds by solving the MWSS problem on each strip, replacing the strips by simple ``gadgets'' and, finally, re-assembling the gadgets to produce a line graph $H$ with the property that any MWSS of $H$ ``corresponds'' to (and can be easily turned into) a MWSS of $G$. Several steps of the algorithm sketched above have a bottleneck time complexity of ${\cal O}(|V||E|)$. In particular, finding the articulation cliques, turning a claw-free graph into a quasi-line graph, solving the MWSS problem in the \{claw, net\}-free strips and in the claw-free strips with stability number not greater than three have that complexity.

\medskip\noindent
In a series of papers we have shown how to perform more efficiently all the bottleneck steps. In particular, in \cite{NobiliSassano17a} an algorithm is described with ${\cal O}(|E| \log |V|)$ time complexity to solve the problem in claw-free graphs with $\alpha(G) \le 3$ and in \cite{NobiliSassano16a} we have proposed a ${\cal O}(|V| \sqrt{|E|})$ algorithm to solve the MWSS problem in \{claw, net\}-free graphs with $\alpha(G) \ge 4$. This final paper addresses the MWSS problem in a claw-free graph $G(V, E)$ by means of a slightly different ungluing operation. We first construct a special maximal stable set $S$ of $G$ which is used to guide the decomposition; second, we perform, directly on the claw-free graph $G$, our ungluing of a proper superset of the family of articulation cliques (\emph{$S$-articulation cliques}) that can be identified more easily and that also produces a decomposition of $G$ into \{claw, net\}-free strips and strips with stability number at most three. Finally we resort to the same procedure proposed by Faenza, Oriolo and Stauffer to solve the MWSS problem in a suitable line graph. The dominant complexity of our algorithm is that of the fastest algorithm to date which solves the MWSS problem in a line graph. In turn, the latter can be derived from any algorithm for the weighted matching problem on a graph $H$ with complexity ${\cal O}(mn \log n)$ (e.g.\ \cite{BD83}, \cite{GMG82}, \cite{Gabow90}), where $m$ is the number of edges and $n$ is the number of nodes of $H$. In fact, if $H$ is the root graph of $G(V, E)$ then $m$ is ${\cal O}(|V|)$ and $n$ (number of cliques in a Krausz partition of $G$) is again ${\cal O}(|V|)$. This implies an ${\cal O}(|V|^2 \log |V|)$ algorithm for the MWSS problem in the line graph $G$. The existence of an algorithm with such a complexity for claw-free graphs was conjectured by Manfred Padberg in 1983.

\section{Notation and Background}

\noindent
For each sub-graph $H$ of a graph $G(V, E)$ we denote by $V(H)$ the set of nodes of $H$ and by $E(H)$ the set of edges of $H$. Moreover, if $W \subseteq V$ we denote $N_G(W)$ (\emph{neighborhood} of $W$ in $G$) the set of nodes in $V \setminus W$ adjacent in $G$ to some node in $W$. If $W = \{w\}$ we simply write $N_G(w)$. We denote by $N_G[W]$ and $N_G[w]$ the sets $N_G(W) \cup W$ and $N_G(w) \cup \{w\}$ (\emph{closed neighborhood} of $W$ and $w$ in $G$). When the graph is unambiguous we simply write $N(W)$ and $N[W]$. If no edge in $E$ has exactly one end-node in $W$ and $W$ is minimal with this property we say that $W$ is (or induces) a connected component of $G$. We also say that two nodes $u$ and $v$ are \emph{distinguished by a subset} $T \subseteq V$ if $u \in T$ and $v \notin T$ or vice versa.

\medskip\noindent
A \emph{clique} is a complete subgraph of $G$ induced by some set of nodes $K \subseteq V$. With a little abuse of notation we also regard the set $K$ as a clique and, for any edge $uv \in E$, both $uv$ and $\{u, v\}$ are said to be a clique. A node $w$ such that $N(w)$ is a clique is said to be \emph{simplicial}. A \emph{claw} is a graph with four nodes $w, x, y, z$ with $w$ adjacent to $x, y, z$ and $x, y, z$ mutually non-adjacent. To highlight its structure, it is denoted as $(w: x, y, z)$. A $P_k$ is a (chordless) path induced by $k$ nodes $u_1, u_2, \dots, u_k$ and will be denoted as $(u_1, \dots, u_k)$. A set $T \subseteq V$ is \emph{complete} (\emph{anticomplete}) to a set $W \subseteq V \setminus T$ if and only if $N(T) \cap W = W$ ($N(T) \cap W = \emptyset$). With a little abuse of notation we regard a node $v \in V$ as the singleton set $T = \{v\}$ and say that $v$ is complete/anticomplete to $W$ ($W$ is complete/anticomplete to $v$). Observe that if $W$ is empty then $T$ is both complete and anticomplete to $W$. A \emph{net} $(x_1, x_2, x_3: y_1, y_2, y_3)$ is a graph induced by a clique $T = \{x_1, x_2, x_3\}$ and three mutually non-adjacent nodes $\{y_1, y_2, y_3\}$ with $N(y_i) \cap T = \{x_i\}$ ($i = 1, 2, 3$). The clique $T$ is said to be a \emph{net triangle}.

\smallskip\noindent
A node $v \in V$ is said to be \emph{regular} if its closed neighborhood can be covered by two (not necessarily distinct) maximal cliques; moreover, if such a cover is unique the node is said to be \emph{strongly regular}. A clique $Q$ is \emph{crucial} for a node $u \in Q$ if $u$ is strongly regular and $Q$ belongs to the unique cover of $N[u]$. A graph $G(V, E)$ is \emph{quasi-line} if all of its nodes are regular. Each line graph is a quasi-line graph and each quasi-line graph is a claw-free graph. A \emph{$5$-wheel} $W_5 = (\bar v: v_1, \dots, v_5)$ is a graph consisting of a chordless cycle $R = (v_1, \dots, v_5)$ called \emph{rim} of $W_5$ and a node $\bar v$ (\emph{hub} of $W_5$) adjacent to every node of $R$. Observe that the hub of $W_5$ is not regular and hence a quasi-line graph does not contain $5$-wheels. 

\smallskip\noindent
Let $S$ be a stable set of a claw-free graph $G(V, E)$. Any node $s \in S$ is said to be \emph{stable}; any node $v \in V \setminus S$ satisfies $|N(v) \cap S| \le 2$ and is called \emph{superfree} if $|N(v) \cap S| = 0$, \emph{free} if $|N(v) \cap S| = 1$ and \emph{bound} if $|N(v) \cap S| = 2$. Observe that, by claw-freeness, a bound node $b$ cannot be adjacent to a node $u \in V \setminus S$ unless $b$ and $u$ have a common neighbor in $S$. For each node $u \in V \setminus S$ we denote by $S(u)$ the set of nodes in $S$ adjacent to $u$. For each $T \subseteq S$ we denote by $F(T)$ the set of free nodes with respect to $S$ which are adjacent to some node $t \in T$. If $T \equiv \{t\}$ we simply write $F(t)$. 

\section{Wings, Similarity and Weakly Normal Cliques}

\noindent
A \emph{bound-wing} defined by $\{s, t\} \subseteq S$ ($s \ne t$) is the set $W^B(s, t) = \{u \in V \setminus S: N(u) \cap S = \{s, t\}\}$. A \emph{free-wing} defined by the ordered pair $(s, t)$ ($s, t \in S$) is the set $W^F(s, t) = \{u \in F(s): N(u) \cap F(t) \ne \emptyset\}$. Observe that, by claw-freeness, any bound node is contained in a single bound-wing. On the other hand, a free node can belong to several free-wings. Moreover, while $W^B(s, t) \equiv W^B(t, s)$, we have $W^F(s, t) \ne W^F(t, s)$ in general (they could both be empty). By slightly generalizing the definition due to Minty \cite{Minty}, we call \emph{wing} defined by $(s, t)$ ($s, t \in S$) the set $W(s, t) = W^B(s, t) \cup W^F(s, t) \cup W^F(t, s)$ if non-empty. We also say that $s$ ($t$) defines $W(s, t)$. Observe that $W(s, t) = W(t, s)$. The nodes $s$ and $t$ are said to be the \emph{extrema} of the wing $W(s, t)$.

\medskip\noindent
Following Schrijver \cite{Schrijver} we say that two nodes $u$ and $v$ in $V \setminus S$ are \emph{similar} ($u \sim v$) if $N(u) \cap S = N(v) \cap S$ and \emph{dissimilar}($u \not\sim v$) otherwise. Clearly, similarity induces an equivalence relation on $V \setminus S$ and a partition in \emph{similarity classes}. Similarity classes can be \emph{bound}, \emph{free} or \emph{superfree} in that they are entirely composed by nodes that are bound, free or superfree with respect to $S$. Bound similarity classes are precisely the bound-wings defined by pairs of nodes of $S$, while each free similarity class contains the (free) nodes adjacent to the same node of $S$. Let $G_F(F(S), E_F)$ be the graph with edge-set $E_F = \{uv \in E: u, v \in F(S), u \not\sim v\}$ (\emph{free dissimilarity graph}). A connected component $D$ of $G_F$ is said to be an \emph{F-clique} defined by $S$ if it induces a maximal clique in $G$. Observe that an F-clique intersects two or more similarity classes; in the first case it is said to be \emph{trivial}. The family of the F-cliques defined by $S$ in $G$ is denoted by ${\cal F}(S)$. Let $Z$ be the set of strongly regular nodes in $S$. For each $s \in Z$, let $(C_s, \bar C_s)$ be the unique pair of maximal cliques covering $N[s]$. The family ${\cal C}(S) = \{(C_s, \bar C_s): s \in Z\}$ is said to be the \emph{$S$-cover} of $G$. With a little abuse of notation, we also say that some clique $C$ belongs to the $S$-cover ${\cal C}(S)$ if $C \in \{C_s, \bar C_s\}$ for some pair $(C_s, \bar C_s) \in {\cal C}(S)$. 

\medskip
\begin{figure}[htbp]
\centering
\begin{tikzpicture}
\GraphInit[vstyle=Normal]

\tikzstyle{VertexStyle}=[shape = circle, fill = black, minimum size = 20pt, text = white, draw]
\Vertex[L=$1$, x=0, y=2]{1}
\Vertex[L=$2$, x=6, y=2]{2}
\Vertex[L=$3$, x=2, y=0]{3}

\tikzstyle{VertexStyle}=[shape = circle, fill = white, minimum size = 20pt, text = black, draw]
\Vertex[L=$4$, x=0, y=0]{4}
\Vertex[L=$6$, x=2, y=2]{6}
\Vertex[L=$7$, x=4, y=2]{7}
\Vertex[L=$8$, x=3, y=1]{8}
\Vertex[L=$9$, x=6, y=0]{9}
\Vertex[L=$5$, x=4, y=0]{5}

\tikzstyle{VertexStyle}=[shape = circle, minimum size = 17pt, draw]
\Vertex[x=2, y=2]{}
\Vertex[x=3, y=1]{}
\Vertex[x=4, y=2]{}
\Vertex[x=6, y=0]{}
\Vertex[x=4, y=0]{}

\Edge(1)(4) \Edge(1)(6)
\Edge(2)(7) \Edge(2)(9)
\Edge(3)(4) \Edge(3)(8) \Edge(3)(5)
\Edge(6)(7) \Edge(6)(8)
\Edge(7)(8) \Edge(7)(9)
\Edge(8)(5)
\Edge(9)(5)

\SetUpEdge[lw = 2pt, color = red]
\Edge(6)(7) \Edge(6)(8) \Edge(7)(8) \Edge(9)(5)

\end{tikzpicture}
\captionsetup{justification=centering, margin=2cm}
\caption{Stable (black), free (double circle) and bound nodes. Wings and F-cliques.}
\label{FigBasicDef}
\end{figure}
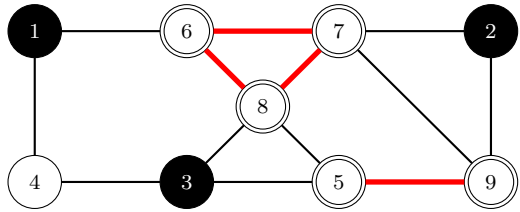

\noindent
\textsl{
\small
In the graph shown in figure~\ref{FigBasicDef}, $S = \{1, 2, 3\}$, $F(S) = \{5, 6, 7, 8, 9\}$ and $3$ defines two wings. Moreover, $W(1, 3) = \{4, 6, 8\}$, $W^B(1, 3) = \{4\}$, $W^F(2, 3) = \{7, 9\}$, $W^F(3, 2) = \{5, 8\}$ and $W(2, 3) = \{5, 7, 8, 9\}$. The dissimilar pairs of free nodes (edges of the free dissimilarity graph) are marked in red, $F_1 = \{6, 7, 8\}$ is a non-trivial F-clique and $F_2 = \{5, 9\}$ is a trivial F-clique. Finally, the unique pair of maximal cliques covering $N[3]$ is $C_{3} = \{3, 4\}$ and $\bar C_{3} = \{3, 5, 8\}$.
}

\medskip\noindent
Two non-adjacent nodes $u, v \in N(Q)$ are said to be \emph{$Q$-distant} if $N(u) \cap N(v) \cap Q = \emptyset$ and \emph{$Q$-close} otherwise. A maximal clique $Q$ is \emph{normal} \cite{LovaszPlummer} if it has three independent neighbors that are mutually $Q$-distant and \emph{weakly normal} if every two non-adjacent nodes in $N(Q)$ are $Q$-distant. 

\begin{Lemma}\label{NormalNetTriangle}
Let $Q$ be a maximal clique in a claw-free graph $G$.
\begin{description}
  \item[\emph{(i)}] $Q$ is normal if and only if it contains a net triangle;
  \item[\emph{(ii)}] if $Q$ is normal and every node in $Q$ is regular then $Q$ is weakly normal;
  \item[\emph{(iii)}] if $Q$ is weakly normal then every node in $Q$ is regular.
\end{description}
\end{Lemma}

\noindent
\emph{Proof}. To prove \emph{(i)} assume first that $Q$ is normal and let $x, y, z$ be three independent and $Q$-distant nodes in $N(Q)$. Let $x', y', z' \in Q$ be neighbors of $x, y, z$, respectively. Then $(x', y', z': x, y, z)$ is a net and $\{x', y', z'\}$ is a net triangle in $Q$. On the other hand, if $(x', y', z': x, y, z)$ is a net with $\{x', y', z'\} \subseteq Q$ we claim that $x, y, z$ are mutually $Q$-distant. Otherwise, without loss of generality, assume that $x$ and $y$ have a common neighbor $q \in Q$. But then $(q: x, y, z')$ is a claw in $G$, a contradiction. Statement \emph{(ii)} have been proved in \cite{LovaszPlummer} (Theorem~12.4.5, Claim~4). Finally, to prove \emph{(iii)} observe that if $Q$ is weakly normal then, for any node $u \in Q$, $Q$ and $N(u) \setminus Q$ are two cliques covering $N[u]$.
\done

\medskip\noindent
\textsl{
\small
In the graph of figure~\ref{FigBasicDef} the clique $\{6, 7, 8\}$ is normal while $\{2, 7, 9\}$ is weakly normal but not normal.
}

\begin{Theorem}\label{Similar}
Let $G(V, E)$ be a claw-free graph and $S$ a maximal stable set of $G$. Then a connected component of $G_F$ intersecting three or more free similarity classes induces a maximal clique in $G$ and hence is a non-trivial F-clique.
\end{Theorem}

\noindent
\emph{Proof}. We first claim that the nodes of any chordless path $P$ in $G_F$ connecting two dissimilar nodes $u, v$ belong only to the similarity classes of $u$ and $v$. In fact, two consecutive nodes of $P$ necessarily belong to different classes. If a node in a third class existed in $P$ we would necessarily have three consecutive nodes $x, y, z$ of $P$ in three different classes. But then $(y: S(y), x, z)$ would be a claw in $G$, a contradiction. Suppose now that a connected component $X$ of $G_F$, intersecting three or more similarity classes, is not a clique in $G$ and let $u, z \in X$ be two non-adjacent nodes in $G$. Suppose first that $S(u)$ and $S(z)$ are two distinct nodes of $S$ and, consequently, that $u \not\sim z$. Let $v \in X$ be a node with $S(v) \notin \{S(u), S(z)\}$, it exists since we assumed that $X$ intersects more than two similarity classes. Let $P_{uv}$ and $P_{vz}$ be chordless paths connecting $u$ to $v$ and, respectively, $v$ to $z$ in $G_F$. By the above claim, $P_{uv}$ contains only nodes in the similarity classes of $u$ and $v$, while $P_{vz}$ contains only nodes in the similarity classes of $v$ and $z$. Let $W_{uz}$ be the walk connecting $u$ to $z$ obtained by chaining $P_{uv}$ and $P_{vz}$ and let $P_{uz}$ be any chordless path connecting $u$ to $z$ whose nodes belong to $W_{uz}$. Since $uz \notin E$, $P_{uz}$ contains at least one node in the similarity class of $v$ and hence contains nodes in three different similarity classes, contradicting the hypothesis that $P_{uz}$ is chordless. It follows that $u$ and $z$ belong to the same similarity class. Moreover, any two dissimilar nodes in $X$ are adjacent in $G$. Let $v \in X$ be a node with $S(v) \ne S(u) \equiv S(z)$. It follows that $uv, vz \in E$ and hence $(v: S(v), u, z)$ is a claw, a contradiction. Hence $X$ is a clique in $G$. To prove that it is also maximal, assume by contradiction that there exists some node $u \in N_G(X)$ complete to $X$. The node $u$ is not free for, otherwise, it would belong to $X$ and is not stable since $X$ intersects more than one similarity class. It follows that $u$ is bound and adjacent to two nodes $s, t \in S$. Moreover, there exists some node $z \in N(X) \cap S$ with $z \ne s, t$. But then, for each node $x \in X \cap N(z)$, $(u: s, t, x)$ is a claw in $G$, a contradiction. The theorem follows.
\done

\begin{Theorem}\label{WeaklyNormalCliques}
Let $G(V, E)$ be a claw-free graph and $S$ a maximal stable set. A weakly normal clique $Q$ of $G$ with $Q \cap S = \emptyset$ belongs to ${\cal F}(S)$.
\end{Theorem}

\noindent
\emph{Proof}. Suppose that there exists a bound node $v \in Q$. Let $W(s, t)$ be the wing containing $v$. Both $s$ and $t$ belong to $N(Q)$ and are adjacent to the node $v \in Q$. But this contradicts the assumption that $Q$ is weakly normal, since $s$ and $t$ are non-adjacent. It follows that $Q$ contains only free nodes, each one of them adjacent to some node in $S$, so $S' = N(Q) \cap S \ne \emptyset$. We have that $S'$ contains at least two nodes, since otherwise the unique node in $S'$ would be complete to $Q$, contradicting its maximality. It follows that $Q$ induces a complete multi-partite subgraph of $G_F$ and hence is contained in some connected component $C$ of $G_F$. If $Q \equiv C$ then it belongs to ${\cal F}(S)$ and the theorem follows. Otherwise there exists a node $x \in C \setminus Q$ adjacent to some dissimilar node $y \in Q$. But then $x$ and $S(y)$ are non-adjacent nodes in $N(y) \setminus Q$, contradicting the assumption that $Q$ is weakly normal.
\done

\section{Ungluing and $S$-articulation cliques}

In \cite{FaenzaOS14} Faenza et al.\ define the concept of \emph{ungluing} of a clique (\emph{partition clique}) into \emph{spikes} and a decomposition operation of a quasi-line graph based on the ungluing of a special family of weakly normal cliques called articulation cliques. A maximal clique $Q$ is an \emph{articulation clique} if it is crucial for each node $u \in Q$. In \cite{FaenzaOS14} (Lemma~3.10) it was shown that in a quasi-line graph a maximal clique containing a net triangle is an articulation clique. In this section we apply a slightly modified decomposition operation (that we keep calling \emph{ungluing}) to the more general class of claw-free graphs and to a different sub-family of weakly normal cliques, properly containing the articulation cliques. To this purpose we first introduce the concept of \emph{canonical stable set}.

\begin{Definition}\label{CanonicalStableSet} (\emph{Canonical stable set})
Let $G(V, E)$ be a connected claw-free graph. A maximal stable set $S$ of $G$ with the property that, for each $s \in S$, $F(s)$ induces a clique in $G$ is said to be \emph{canonical}.
\done
\end{Definition}

\medskip\noindent
The stable set $S = \{v_1, v_2, v_3\}$ in figure~\ref{FigBasicDef} is canonical. In Section~\ref{Finding} we will show that a canonical stable set always exists and can be efficiently found. Hence, in what follows we assume that a connected claw-free graph $G(V, E)$ with $\alpha(G) \ge 4$ and a canonical stable set $S$ of $G$ are given. We let ${\cal C} \equiv {\cal C}(S)$ be the $S$-cover of $G$ and ${\cal F} \equiv {\cal F}(S)$ the family of F-cliques.

\begin{Definition}\label{SArticulationCliques} (\emph{$S$-articulation clique})
The family of \emph{$S$-articulation cliques} ${\cal S}$ is obtained from ${\cal C} \cup {\cal F}$ by removing:
\begin{description}
  \item[\emph{(i)}] any clique $Q$ which is not weakly normal;
  \item[\emph{(ii)}] any pair of weakly normal cliques $Q, K$ such that $N(Q \cap K) \not\subseteq (Q \cup K)$.
\done
\end{description}
\end{Definition}

\medskip\noindent
We now introduce the concepts of rigid (soft) edges and rigid structure and describe our new definitions of ungluing, spike and strip.

\begin{Definition}\label{SoftEdge} (\emph{Soft and rigid edges, rigid structure})
An edge $uv \in E$ is \emph{soft} if $u$ and $v$ are distinguished by some clique $Q \in {\cal S}$ or $N(u) \cap N(v)$ is a clique in $G$. An edge which is not soft is said to be \emph{rigid} in $G$. Let $E_R \subseteq E$ be the set of rigid edges of $G$. The graph $G_R(V, E_R)$ is said to be the \emph{rigid structure} of $G$ and an induced subgraph $G[U]$ ($U \subseteq V$) is said to be \emph{rigid in $G$} if $G_R[U]$ is connected, \emph{soft in $G$} otherwise.
\done
\end{Definition}

\begin{Definition}\label{Ungluing} (\emph{Ungluing, spike, strip})
Let $G_R$ be the rigid structure of $G$. The \emph{ungluing} of $G$ is the graph $G_{\cal S}(V, E_{\cal S})$ obtained by removing any edge $uv$ belonging to some $Q \in {\cal S}$ and such that $u$ and $v$ belong to different components of $G_R$. For each clique $Q \in {\cal S}$, any connected component $K \subseteq Q$ of $G_{\cal S}[Q]$ is said to be a \emph{spike} of $Q$. We denote by ${\cal K}$ the family of all the spikes and call \emph{strip} a connected component of $G_{\cal S}$ containing at most two spikes. 
\done
\end{Definition}

\medskip\noindent
The main result of this section is the following.

\begin{Theorem}\label{S-StripTheorem}
Every connected component of $G_{\cal S}$ is a strip.
\done
\end{Theorem}

\medskip\noindent
The rest of the section is devoted to the proof of the above theorem. In particular, we first prove some properties of spikes and $S$-articulation cliques; then we introduce a new graph $G^+$ and prove that such a graph is claw-free and that its regular connected components are net-free; moreover, we show that every irregular connected component $C$ of $G_{\cal S}$ is a strip with $\alpha(C) \le 3$ and, finally, that each regular connected component of $G_{\cal S}$ is a strip.

\begin{Lemma}\label{ArticulationSArticulation}
The family of articulation cliques of $G$ is contained in ${\cal S}$.
\end{Lemma}

\noindent
\emph{Proof.} Let $Q$ be an articulation clique of $G$. $Q$ is weakly normal since, otherwise, a node $u \in Q$ would exist with $N(u) \setminus Q$ not a clique. Moreover, by Theorem~\ref{WeaklyNormalCliques}, $Q$ is contained either in ${\cal F}$ or contains some stable node $s \in S$. In the latter case, since $Q$ is an articulation clique we have that $s$ is strongly regular and hence $Q$ belongs to ${\cal C}$. If $Q$ does not belong to ${\cal S}$, then there exists a weakly normal clique $K$ with the property that $N(Q \cap K) \not\subseteq Q \cup K$. Let $w$  be a node in $Q \cap K$ with $N(w) \not\subseteq Q \cup K$. Let $\bar Q$ be a maximal clique containing $N(w) \setminus Q$ and $\bar K$ be a maximal clique containing $N(w) \setminus K$. Then we have $N(w) \subseteq Q \cup \bar Q$ and $N(w) \subseteq K \cup \bar K$ with $\bar Q \ne K$ and $\bar K \ne Q$, contradicting the assumption that $Q$ is an articulation clique.
\done

\begin{Lemma}\label{SpikeInQ}
A spike $K \in {\cal K}$ intersecting some clique $Q \in {\cal S}$ is contained in $Q$.
\end{Lemma}

\noindent
\emph{Proof.} Assume that there exist nodes $u, v \in K$ with $u \in Q$ and $v \notin Q$. Since $G_R[K]$ is connected, there exists in $G[K]$ a path $P = (u \equiv v_1, \dots, v_i \equiv v)$ composed of rigid edges. Let $v_j$ be the first node of $P$ not in $Q$ (possibly $v_j \equiv v_i$). We have that the nodes $v_{j-1}, v_j$ are distinguished by $Q$, so the edge $v_{j-1} v_j$ is soft, a contradiction.
\done

\medskip\noindent
The spikes produced by the ungluing of articulation cliques in quasi-line graphs as described by Faenza et al.\ (\cite{FaenzaOS14}) are disjoint cliques in $G$. Our extended definition of ungluing preserves this property.

\begin{Lemma}\label{SpikesAreDisjoint}
Two spikes in ${\cal K}$ have empty intersection.
\end{Lemma}

\noindent
\emph{Proof.} Let $K_i$ be a spike of $Q_i \in {\cal S}$ ($i = 1, 2$) and assume, by contradiction, that $K_1$ and $K_2$ are distinct and have non-empty intersection. By Lemma~\ref{SpikeInQ} $Q_1$ contains $K_2$. But then $K_1$ and $K_2$ are both spikes of $Q_1$ with non-empty intersection, a contradiction.
\done

\medskip
\begin{figure}[htbp]
\centering
\begin{tikzpicture}
\GraphInit[vstyle=Normal]

\tikzstyle{VertexStyle}=[shape = circle, fill = black, minimum size = 20pt, text = white, draw]
\Vertex[L=$3$, x=2, y=2]{3}
\Vertex[L=$8$, x=6, y=0]{8}
\Vertex[L=$10$, x=8, y=2]{10}

\tikzstyle{VertexStyle}=[shape = circle, fill = white, minimum size = 20pt, text = black, draw]
\Vertex[L=$1$, x=0, y=2]{1}
\Vertex[L=$2$, x=0, y=0]{2}
\Vertex[L=$4$, x=2, y=0]{4}
\Vertex[L=$5$, x=4, y=2]{5}
\Vertex[L=$6$, x=4, y=0.5]{6}
\Vertex[L=$7$, x=6, y=2]{7}
\Vertex[L=$9$, x=7, y=1]{9}
\Vertex[L=$11$, x=8, y=0]{11}
\Vertex[L=$12$, x=9 ,y=1]{12}

\draw [blue] plot [smooth cycle] coordinates {(0,2.5) (-0.5,2) (-0.5,0) (0,-0.5) (2.5,2) (2,2.5)}; 
\draw [blue] plot [smooth cycle] coordinates {(0,2.5) (-0.5,2) (0,1.5) (4,0) (4.5,0.5) (4,1)}; 
\draw [red] plot [smooth cycle] coordinates {(2,2.5) (1.5,2) (1.5,0) (2,-0.5) (4.5,2) (4,2.5)}; 
\draw [blue] plot [smooth cycle] coordinates {(2,-0.5) (1.5,0) (2,0.5) (3.5,2) (4,2.5) (6,2.5) (6.5, 2) (6.5,0) (6,-0.5) }; 
\draw [blue] plot [smooth cycle] coordinates {(6,-0.5) (5.5,0) (7,1.5) (8.5,0) (8,-0.5)}; 
\draw [blue] plot [smooth cycle] coordinates {(5.5,2) (6,2.5) (8,2.5) (8.5,2) (7,0.5)}; 
\draw [blue] plot [smooth cycle] coordinates {(7.5,2) (8,2.5) (9.5,1) (8,-0.5) (7.5,0)}; 

\tikzstyle{VertexStyle}=[shape = circle, minimum size = 17pt, draw]
\Vertex[L=$1$, x=0, y=2]{}
\Vertex[L=$2$, x=0, y=0]{}
\Vertex[L=$6$, x=4, y=0.5]{}
\Vertex[L=$12$, x=9, y=1]{}

\Edge(1)(2) \Edge(1)(3) \Edge(1)(6)
\Edge(2)(3) \Edge(2)(4)
\Edge(3)(4) \Edge(3)(5)
\Edge(4)(5) \Edge(4)(6) \Edge(4)(7) \Edge(4)(8)
\Edge(5)(6) \Edge(5)(7) \Edge(5)(8)
\Edge(6)(7) \Edge(6)(8)
\Edge(7)(8) \Edge(7)(9) \Edge(7)(10)
\Edge(8)(9) \Edge(8)(11)
\Edge(9)(10) \Edge(9)(11)
\Edge(10)(11) \Edge(10)(12)
\Edge(11)(12)

\SetUpEdge[lw = 2pt, color = red]
\Edge(2)(3) \Edge(4)(5)

\end{tikzpicture}
\captionsetup{justification=centering, margin=1.5cm}
\caption{Articulation cliques, $S$-articulation cliques and rigid edges}
\label{FigArticulationUngluing}
\end{figure}
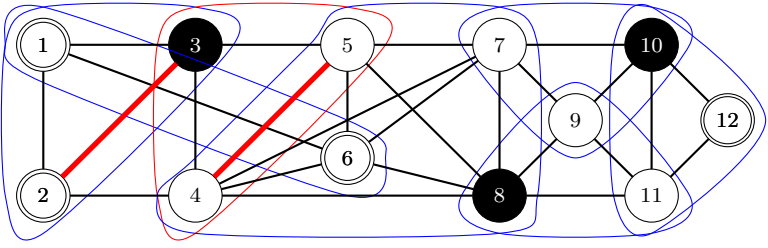

\medskip
\begin{figure}[htbp]
\centering
\begin{tikzpicture}
\GraphInit[vstyle=Normal]

\tikzstyle{VertexStyle}=[shape = circle, fill = black, minimum size = 20pt, text = white, draw]
\Vertex[L=$3$, x=2, y=2]{3}
\Vertex[L=$8$, x=6, y=0]{8}
\Vertex[L=$10$, x=8, y=2]{10}

\tikzstyle{VertexStyle}=[shape = circle, fill = white, minimum size = 20pt, text = black, draw]
\Vertex[L=$1$, x=0, y=2]{1}
\Vertex[L=$2$, x=0, y=0]{2}
\Vertex[L=$4$, x=2, y=0]{4}
\Vertex[L=$5$, x=4, y=2]{5}
\Vertex[L=$6$, x=4, y=0.5]{6}
\Vertex[L=$7$, x=6, y=2]{7}
\Vertex[L=$9$, x=7, y=1]{9}
\Vertex[L=$11$, x=8, y=0]{11}
\Vertex[L=$12$, x=9, y=1]{12}

\tikzstyle{VertexStyle}=[shape = circle, minimum size = 17pt, draw]
\Vertex[L=$1$, x=0, y=2]{}
\Vertex[L=$2$, x=0, y=0]{}
\Vertex[L=$6$, x=4, y=0.5]{}
\Vertex[L=$12$, x=9, y=1]{}

\Edge(2)(3) \Edge(2)(4)
\Edge(3)(4) \Edge(3)(5)
\Edge(4)(5)
\Edge(7)(8) \Edge(7)(9) \Edge(7)(10)
\Edge(8)(9) \Edge(8)(11)
\Edge(9)(10) \Edge(9)(11)
\Edge(10)(11)

\SetUpEdge[style=dotted]
\Edge(1)(2) \Edge(1)(3) \Edge(1)(6)
\Edge(4)(6) \Edge(4)(7) \Edge(4)(8)
\Edge(5)(6) \Edge(5)(7) \Edge(5)(8)
\Edge(6)(7) \Edge(6)(8)
\Edge(10)(12)
\Edge(11)(12)

\end{tikzpicture}
\captionsetup{justification=centering, margin=2cm}
\caption{Ungluing of articulation cliques}
\label{FigUngluingFaenza}
\end{figure}
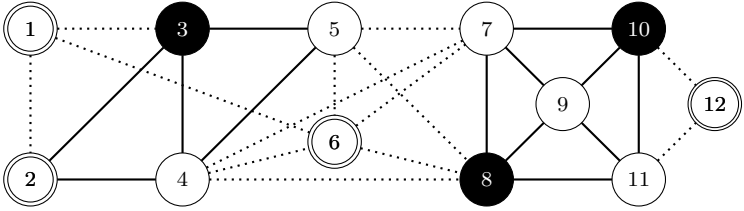

\medskip
\begin{figure}[htbp]
\centering
\begin{tikzpicture}
\GraphInit[vstyle=Normal]

\tikzstyle{VertexStyle}=[shape = circle, fill = black, minimum size = 20pt, text = white, draw]
\Vertex[L=$3$, x=2, y=2]{3}
\Vertex[L=$8$, x=6, y=0]{8}
\Vertex[L=$10$, x=8, y=2]{10}

\tikzstyle{VertexStyle}=[shape = circle, fill = white, minimum size = 20pt, text = black, draw]
\Vertex[L=$1$, x=0, y=2]{1}
\Vertex[L=$2$, x=0, y=0]{2}
\Vertex[L=$4$, x=2, y=0]{4}
\Vertex[L=$5$, x=4, y=2]{5}
\Vertex[L=$6$, x=4, y=0.5]{6}
\Vertex[L=$7$, x=6, y=2]{7}
\Vertex[L=$9$, x=7, y=1]{9}
\Vertex[L=$11$, x=8, y=0]{11}
\Vertex[L=$12$, x=9, y=1]{12}

\tikzstyle{VertexStyle}=[shape = circle, minimum size = 17pt, draw]
\Vertex[L=$1$, x=0, y=2]{}
\Vertex[L=$2$, x=0, y=0]{}
\Vertex[L=$6$, x=4, y=0.5]{}
\Vertex[L=$12$, x=9, y=1]{}

\Edge(2)(3) \Edge(2)(4)
\Edge(3)(4) \Edge(3)(5)
\Edge(4)(5)

\SetUpEdge[style=dotted]
\Edge(1)(2) \Edge(1)(3) \Edge(1)(6)
\Edge(4)(6) \Edge(4)(7) \Edge(4)(8)
\Edge(5)(6) \Edge(5)(7) \Edge(5)(8)
\Edge(6)(7) \Edge(6)(8)
\Edge(7)(8) \Edge(7)(9) \Edge(7)(10)
\Edge(8)(9) \Edge(8)(11)
\Edge(9)(10) \Edge(9)(11)
\Edge(10)(11) \Edge(10)(12)
\Edge(11)(12)

\end{tikzpicture}
\captionsetup{justification=centering, margin=2cm}
\caption{Ungluing of $S$-articulation cliques}
\label{FigUngluingNS}
\end{figure}
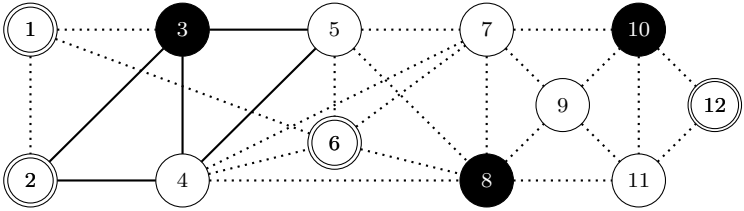

\noindent
\textsl{
\small
The graph $G$ in figure~\ref{FigArticulationUngluing} contains the articulation cliques $\{4, 5, 6, 7, 8\}$, $\{1, 2, 3\}$, $\{1, 6\}$, $\{10, 11, 12\}$. The canonical stable set $S = \{3, 8, 10\}$ defines the trivial $F$-clique $\{1, 6\}$ and the cliques of ${\cal C}$: $C_3 = \{1, 2, 3\}$, $\bar C_3 = \{3, 4, 5\}$, $C_8 = \{4, 5, 6, 7, 8\}$, $\bar C_8 = \{8, 9,11\}$, $C_{10} = \{7, 9, 10\}$, $\bar C_{10} = \{10, 11, 12\}$. The clique $\bar C_3$ (red) is not weakly normal and hence does not belong to the family ${\cal S}$ of $S$-articulation cliques. All the other cliques of ${\cal F} \cup {\cal C}$ are $S$-articulation cliques. The rigid edges are marked in red. The graph in figure~\ref{FigUngluingFaenza} is the ungluing of $G$ as defined in \cite{FaenzaOS14} with respect to the articulation cliques. The graph in figure~\ref{FigUngluingNS} is the ungluing of $G$ as defined in this paper. Note that the two ungluing operations applied to the same clique $\{10, 11, 12\}$ produce different results. 
}

\begin{Lemma}\label{NoThreeSArticulation}
If a node $v \in V$ belongs to more than two distinct weakly normal cliques in ${\cal F} \cup {\cal C}$ then none of them is a $S$-articulation clique.
\end{Lemma}

\noindent
\emph{Proof.} Assume by contradiction that there exist three distinct weakly normal cliques $Q_1, Q_2, Q_3 \in {\cal F} \cup {\cal C}$ with $v \in Q_1 \cap Q_2 \cap Q_3$ and $Q_3 \in {\cal S}$. Since $Q_1$ and $Q_2$ are distinct and maximal, there exist nodes $x \in Q_1 \setminus Q_2$ and $y \in Q_2 \setminus Q_1$ which are non-adjacent, with $\{x, y\} \subseteq N(v)$. The nodes $x$ and $y$ cannot both belong to the clique $Q_3$; hence, without loss of generality, we can assume $y \in Q_2 \setminus (Q_1 \cup Q_3)$. But then we have $v \in Q_1 \cap Q_3$ and $N(v) \not\subseteq Q_1 \cup Q_3$, contradicting the assumption that $Q_3$ belongs to ${\cal S}$.
\done

\begin{Definition}\label{Lifting} (\emph{Lifting and lifting nodes, extension})
Let $G_{\cal S}$ be the ungluing of $G$ and ${\cal K}$ be the family of spikes defined by ${\cal S}$. The graph $G^+(V^+, E^+)$ obtained from $G_{\cal S}$ by adding the nodes $L = \{ q_{K} : K \in {\cal K}\}$ and making each $q_{K}$ complete to $K$ is said to be the \emph{lifting} of $G$ and the nodes in $L$ are said to be the \emph{lifting nodes} of $G^+$. Finally, the stable set $S^+ = S \cup (L \setminus N(S))$ of $G^+$ is said to be the \emph{extension} of $S$ in $G^+$.
\done  
\end{Definition}

\medskip\noindent
Observe that, by construction, the lifting nodes are simplicial in $G^+$. In what follows we will denote by $N^+(K)$ ($N^+[K]$) the neighborhood (closed neighborhood) of some set $K$ in $G^+$. If $K$ is the singleton $u$ we will simply write $N^+(u)$ ($N^+[u]$). 

\medskip
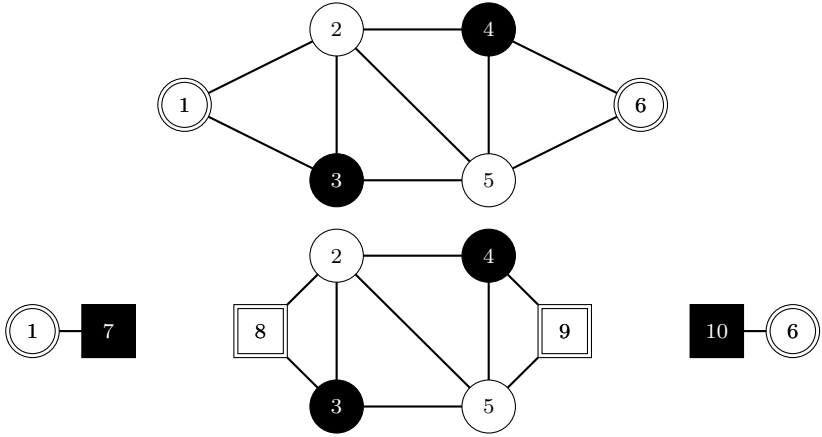
\begin{figure}[htbp]
\centering
\begin{tikzpicture}
\GraphInit[vstyle=Normal]

\tikzstyle{VertexStyle}=[shape = circle, fill = black, minimum size = 20pt, text = white, draw]
\Vertex[L=$3$, x=4, y=0]{3}
\Vertex[L=$4$, x=6, y=2]{4}

\tikzstyle{VertexStyle}=[shape = rectangle, fill = black, minimum size = 20pt, text = white, draw]
\Vertex[L=$7$, x=1, y=1]{7}
\Vertex[L=$10$, x=9, y=1]{10}

\tikzstyle{VertexStyle}=[shape = circle, fill = white, minimum size = 20pt, text = black, draw]
\Vertex[L=$1$, x=0, y=1]{1}
\Vertex[L=$2$, x=4, y=2]{2}
\Vertex[L=$5$, x=6, y=0]{5}
\Vertex[L=$6$, x=10, y=1]{6}

\tikzstyle{VertexStyle}=[shape = rectangle, fill = white, minimum size = 20pt, text = black, draw]
\Vertex[L=$8$, x=3, y=1]{8}
\Vertex[L=$9$, x=7, y=1]{9}

\tikzstyle{VertexStyle}=[shape = circle, minimum size = 17pt, draw]
\Vertex[L=$1$, x=0, y=1]{}
\Vertex[L=$6$, x=10, y=1]{}

\tikzstyle{VertexStyle}=[shape = rectangle, minimum size = 17pt, draw]
\Vertex[L=$8$, x=3, y=1]{}
\Vertex[L=$9$, x=7, y=1]{}

\Edge(1)(7)
\Edge(2)(3) \Edge(2)(4) \Edge(2)(5) \Edge(2)(8)
\Edge(3)(5) \Edge(3)(8)
\Edge(4)(5) \Edge(4)(9)
\Edge(5)(9)
\Edge(6)(10)

\tikzstyle{VertexStyle}=[shape = circle, fill = black, minimum size = 20pt, text = white, draw]
\Vertex[L=$3$, x=4, y=3]{13}
\Vertex[L=$4$, x=6, y=5]{14}

\tikzstyle{VertexStyle}=[shape = circle, fill = white, minimum size = 20pt, text = black, draw]
\Vertex[L=$1$, x=2, y=4]{11}
\Vertex[L=$2$, x=4, y=5]{12}
\Vertex[L=$5$, x=6, y=3]{15}
\Vertex[L=$6$, x=8, y=4]{16}

\tikzstyle{VertexStyle}=[shape = circle, minimum size = 17pt, draw]
\Vertex[L=$1$, x=2, y=4]{}
\Vertex[L=$6$, x=8, y=4]{}

\Edge(11)(12) \Edge(11)(13)
\Edge(12)(13) \Edge(12)(14) \Edge(12)(15)
\Edge(13)(15)
\Edge(14)(15) \Edge(14)(16)
\Edge(15)(16)

\end{tikzpicture}
\captionsetup{justification=centering, margin=2cm}
\caption{A claw-free graph $G$ with a stable set $S$ and their lifting and extension}
\label{FigLifting}
\end{figure}

\medskip\noindent
\textsl{
\small
In the first graph $G$ shown in figure~\ref{FigLifting}, the stable set $S = \{3, 4\}$ is canonical. The family ${\cal S}$ contains the cliques $C_3 = \{1, 2, 3\}$ and $C_4 = \{4, 5, 6\}$. The family ${\cal K}$ contains the spikes $K_7 = \{1\}$, $K_8 = \{2, 3\}$, $K_9 = \{4, 5\}$ and $K_{10} = \{6\}$. The second graph is the lifting $G^+$ of $G$ with lifting nodes $L = \{7, 8, 9, 10\}$. The stable set $S^+ = \{3, 4, 7, 10\}$ is the extension of $S$.
}

\begin{Lemma}\label{ThreeEdges}
If $wu$ and $wv$ are edges in $E_{\cal S}$ while $uv \notin E_{\cal S}$ then $uv$ does not belong to $E$.
\end{Lemma}

\noindent
\emph{Proof.} Assume, by contradiction, that $uv$ belongs to $E \setminus E_{\cal S}$. It follows that the edge $uv$ is soft and there exists some (maximal) clique $Q \in {\cal S}$ containing $u$ and $v$ in different spikes. The node $w$ does not belong to $Q$ since, otherwise, either $wu$ or $wv$ would not belong to $E_{\cal S}$. Hence, there exists some node $\bar w \in Q$ with $w \bar w \notin E$. It follows that $N(u) \cap N(v)$ is not a clique in $G$ and, by Definition~\ref{SoftEdge}, $u$ and $v$ are distinguished by some clique $Q_h \in {\cal S}$. Without loss of generality, we can assume $u \in Q_h$ and $v \notin Q_h$. The node $w$ belonging to $N(Q \cap Q_h)$ must belong to $Q_h$ since, otherwise, $N(Q \cap Q_h) \not\subseteq Q \cup Q_h$ and the assumption that both $Q$ and $Q_h$ are $S$-articulation cliques would be contradicted. But then the nodes $u$ and $w$ belong to the same spike $K \subseteq Q_h$ with $K \cap Q \ne \emptyset$ and $K \not\subseteq Q$, contradicting Lemma~\ref{SpikeInQ}. The lemma follows.
\done

\begin{Lemma}\label{LiftingClawFree}
The graph $G^+$ is claw-free.
\end{Lemma}

\noindent
\emph{Proof.} Assume, by contradiction, that there exists a claw $(w: x, y, z)$ in $G^+$. Observe that $w$ is not a lifting node (since it is not simplicial). Moreover, by Lemma~\ref{SpikesAreDisjoint}, the node $w$ belongs to at most one spike and hence $|\{x, y, z\} \cap L| \le 1$. If $\{x, y, z\} \cap L = \emptyset$ we have, By Lemma~\ref{ThreeEdges}, $xy, yz, xz \notin E$ and hence $(w: x, y, z)$ is a claw in $G$, a contradiction. If, on the other hand, $|\{x, y, z\} \cap L| = 1$ then, without loss of generality, we can assume $z \in L$. Let $K$ be the spike complete to $z$ and let $Q \in {\cal S}$ be a $S$-articulation clique containing $K$. The nodes $x$ and $y$ do not belong to $K$ since they are non-adjacent to $z$ and do not belong to a different spike of $Q$ since they are adjacent to $w$ in $G_{\cal S}$. It follows that they belong to $N_G(Q)$. Moreover, by Lemma~\ref{ThreeEdges}, we have $xy \notin E$. But then $w \in Q$ is a common neighbor of two non-adjacent nodes in $N_G(Q)$, contradicting the assumption that $Q$ is weakly normal in $G$. The theorem follows.
\done

\medskip\noindent
Every regular connected component of $G^+$ is a quasi-line graph. The following lemma shows that it is also net-free. 

\begin{Lemma}\label{LiftingNetFree}
Every regular connected component $C$ of $G^+$ is net-free.
\done
\end{Lemma}

\medskip\noindent
To prove the above lemma we first need some technical results.

\begin{Lemma}\label{PreservingSArticulation0}
Let $Q$ be a maximal clique in $G$. If $Q$ is a normal clique in $G^+$ contained in a regular connected component then it is a $S$-articulation clique in $G$. 
\end{Lemma}

\noindent
\emph{Proof.} Since $Q$ is normal in $G^+$, by \emph{(i)} of Lemma~\ref{NormalNetTriangle} there exists a net  $(x_1, x_2, x_3: y_1, y_2, y_3)$ in $G^+$ with $x_1, x_2, x_3 \in Q$. Moreover, by \emph{(ii)} of Lemma~\ref{NormalNetTriangle}, $Q$ is also weakly normal in $G^+$ and, by \cite{FaenzaOS14} (Lemma~3.10), it is an articulation clique in $G^+$.

\medskip\noindent
\emph{Claim (i). The clique $Q$ is weakly normal in $G$.}

\smallskip\noindent
\emph{Proof.} Assume, by contradiction, that there exist non-adjacent nodes $v_1, v_2 \in N_G(Q)$ both adjacent to some node $u \in Q$. If $v_1 u \in E \setminus E^+$ let $\bar v_1 \in L$ be the lifting node adjacent to $u$ in $G^+$, otherwise let $\bar v_1 \equiv v_1$. Analogously define $\bar v_2$. Observe that $\bar v_1 \ne \bar v_2$. This is clear if $\bar v_1 \equiv v_1$ or $\bar v_2 \equiv v_2$. If both $\bar v_1$ and $\bar v_2$ belong to $L$ they are distinct since $v_1$ and $v_2$ are not adjacent in $G$ and hence belong to different spikes. In any case $\bar v_1$ and $\bar v_2$ are non-adjacent nodes in $N^+(u) \setminus Q$, contradicting the assumption that $Q$ is weakly normal in $G^+$. The claim follows.

\noindent
\emph{End of Claim (i).}

\medskip\noindent
\emph{Claim (ii). For each node $u \in Q$ which does not belong to a $S$-articulation clique in $G$ we have that $Q$ is \emph{crucial} for $u$ in $G$.}

\smallskip\noindent
\emph{Proof.} If $u$ is simplicial then $Q$ is trivially crucial for $u$, hence assume that $u$ is not simplicial. First, we show that if $Q$ is not crucial for $u$ then it cannot belong to all the pairs of maximal cliques covering $N_G[u]$. Otherwise, there exist in $G$ maximal cliques $Q' \ne Q''$ such that $N_G[u] \subseteq (Q \cup Q')$ and $N_G[u] \subseteq (Q \cup Q'')$. Since $Q' \ne Q''$ are maximal cliques, it follows that both contain $u$ and there exist non-adjacent nodes $v' \in Q' \setminus Q''$ and $v'' \in Q'' \setminus Q'$. But then $v' \in N_G[u] \setminus Q''$ and $v'' \in N_G[u] \setminus Q'$ must both belong to $Q$, a contradiction. Hence, if $Q$ is not crucial for $u$, there exist in $G$ maximal cliques $Q', Q'' \ne Q$ such that $N_G[u] \subseteq (Q' \cup Q'')$. Since $Q$ is an articulation clique in $G^+$ then either $Q'$ or $Q''$ (say $Q'$) is not a clique in $G^+$. Hence $Q'$ contains two nodes $v$ and $z$ which belong to different spikes of a $S$-articulation clique $\bar Q$. But then, by Lemma~\ref{ThreeEdges}, $u v \in E \setminus E^+$ or $u z \in E \setminus E^+$, contradicting the assumption that $u$ does not belong to a $S$-articulation clique. The claim follows.

\noindent
\emph{End of Claim (ii).}

\medskip\noindent
By Claim~\emph{(i)} and Theorem~\ref{WeaklyNormalCliques}, $Q$ is either contained in ${\cal F}$ or contains some stable node $s \in S$. In this latter case, we claim that $s$ is strongly regular and hence, by definition of $S$-cover, $Q$ belongs to ${\cal C}$. In fact, if $s$ belongs to some $S$-articulation clique in $G$ then, by definition, $s$ is strongly regular; on the other hand, if $s$ does not belong to a $S$-articulation clique then, by Claim~\emph{(ii)}, $Q$ is crucial for $s$ in $G$ and, again, $s$ is strongly regular. It follows that $Q$ belongs to ${\cal F} \cup {\cal C}$. 

\medskip\noindent
Assume that $Q$ is not a $S$-articulation clique. Hence there exists, by Definition~\ref{SArticulationCliques}, a weakly normal clique $\bar Q \in {\cal F} \cup {\cal C}$ such that $N_G(\bar Q \cap Q) \not\subseteq (\bar Q \cup Q)$. Observe that also $\bar Q$ is not a $S$-articulation clique. Moreover, no $S$-articulation clique $Q_i$ contains any node $u \in \bar Q \cap Q$ for, otherwise, the node $u$ would belong to three distinct weakly normal cliques in ${\cal F} \cup {\cal C}$, namely $Q$, $\bar Q$ and $Q_i$, with $Q_i \in {\cal S}$, contradicting Lemma~\ref{NoThreeSArticulation}. As a consequence, no node in $L$ is adjacent to some node $u \in \bar Q \cap Q$ and any edge $uv \in E$ with $u \in \bar Q \cap Q$ also belongs to $E^+$.

\medskip\noindent
Let $v$ be any node in $N_G(\bar Q \cap Q) \setminus (\bar Q \cup Q)$ and $u$ a node in $N_G(v) \cap (\bar Q \cap Q)$. It follows that $uv \in E^+$. Since $\bar Q$ is weakly normal in $G$, we have that $v$ is complete to $Q \setminus \bar Q$ in $G$. Moreover, by Lemma~\ref{ThreeEdges}, $v$ is also complete to $Q \setminus \bar Q$ in $G^+$. Suppose now that $x_1, x_2 \notin \bar Q$. It follows that $v x_1, v x_2 \in E^+$ and, since $Q$ is weakly normal in $G^+$, that $v y_1, v y_2 \in E^+$. If $x_3 \notin \bar Q$ we have that $v x_3 \in E^+$ and $v y_3 \in E^+$. But then $(v: y_1, y_2, y_3)$ is a claw in $G^+$ contradicting Lemma~\ref{LiftingClawFree}. It follows that $x_3$ belongs to $\bar Q \cap Q$ and hence $y_3 \notin L$. Observe that, by Lemma~\ref{ThreeEdges}, the edge $x_2 y_3$ does not belong to $E$. If $y_3 \notin \bar Q$ then the non-adjacent nodes $x_2, y_3 \in N_G(\bar Q)$ have the common neighbor $x_3 \in \bar Q$, contradicting the hypothesis that $\bar Q$ is weakly normal in $G$. It follows that $y_3$ belongs to $\bar Q$ and hence is adjacent to $u$. Moreover, since $Q$ is weakly normal in $G$ (Claim~\emph{(i)}), the nodes $v, y_3 \in N_G(u)$ must be adjacent. But then $(v: y_1, y_2, y_3)$ is a claw in $G^+$ contradicting Lemma~\ref{LiftingClawFree}. 

\smallskip\noindent
So we can assume, without loss of generality, that $x_1$ and $x_2$ belong to $\bar Q \cap Q$ and hence $y_1, y_2 \notin L$. Observe that, by Lemma~\ref{ThreeEdges}, the edges $y_1 x_2$ and $y_1 x_3$ do not belong to $E$ and, consequently, $y_1$ does not belong to $\bar Q$. It follows that $x_3 \in \bar Q$ for, otherwise, the non-adjacent nodes $x_3, y_1 \in N_G(\bar Q)$ would have the common neighbor $x_1$ in $\bar Q$, contradicting the hypothesis that $\bar Q$ is weakly normal in $G$. But then also $y_3$ does not belong to $L$. Consequently, every node $y \in \bar Q \setminus Q$ must be adjacent in $G$ to $y_1, y_2$ and $y_3$ since $Q$ is weakly normal in $G$. But then $(y: y_1, y_2, y_3)$ is a claw in $G$, a contradiction. We can conclude that $Q$ is a $S$-articulation clique and the lemma follows.
\done

\begin{Lemma}\label{PreservingSArticulation}
Let $\{x_1, x_2, x_3\}$ be a net triangle in a regular connected component of $G^+$. Then there exists a $S$-articulation clique $Q \in {\cal S}$ containing $\{x_1, x_2, x_3\}$.
\end{Lemma}

\noindent
\emph{Proof.} Let $(x_1, x_2, x_3: y_1, y_2, y_3)$ be a net in $G^+$ and let $Q'$ be a maximal clique in $G^+$ containing $\{x_1, x_2, x_3\}$. If two nodes in $\{y_1, y_2, y_3\}$, say $y_1$ and $y_2$, had a common neighbor $u \in Q'$ we would have the claw $(u: x_3, y_1, y_2)$ in $G^+$, a contradiction. It follows that $y_1, y_2, y_3$ are mutually $Q'$-distant in $G^+$ and hence $Q'$ is normal in $G^+$. Moreover, by hypothesis, $Q'$does not contain irregular nodes. If $Q'$ is also a maximal clique in $G$ then, by Lemma~\ref{PreservingSArticulation0}, $Q \equiv Q'$ is a $S$-articulation clique and we are done. Hence we can assume that $Q'$ is not a maximal clique in $G$ and that no $S$-articulation clique in $G$ contains $\{x_1, x_2, x_3\}$. Let $\bar Q$ be a maximal clique in $G_{\cal S}$ containing  $Q' \setminus L$ and observe that $\bar Q$ is a clique in $G$ and contains $\{x_1, x_2, x_3\}$. In fact, $x_1, x_2, x_3 \notin L$ since they are not simplicial in $G^+$. Moreover, $\bar Q$ is not maximal in $G$ (otherwise $\bar Q \equiv Q'$ and $Q'$ would be maximal in $G$). Let $Q$ be a maximal clique in $G$ containing $\bar Q$ (observe that $Q \setminus \bar Q \ne \emptyset$). By assumption $Q$ is not a $S$-articulation clique. Let $F$ be the set of edges $uv$ with $u \in \bar Q$ and $v \in Q \setminus \bar Q$. For each node $v \in Q \setminus \bar Q$ at least one edge $uv \in F$ belongs to $E \setminus E^+$, otherwise $v$ would belong to $\bar Q$. If some edge $\bar u v \in F$ belonged to $E^+$ then the three nodes $u$, $\bar u$ and $v$ would violate Lemma~\ref{ThreeEdges}. As a consequence the set $F$ is contained in $E \setminus E^+$ and hence each edge $uv \in F$ (and each node $u \in \bar Q$) belongs to some $S$-articulation clique. 

\medskip\noindent
Assume that there exists some node $u \in \bar Q$ belonging to two $S$-articulation cliques $Q_1$ and $Q_2$. By assumption $\bar Q \not\subseteq Q_1$, so let $v$ be some node in $\bar Q \setminus Q_1$. Observe that $v$ belongs to $N_G(Q_1 \cap Q_2)$. If $v$ does not belong to $Q_2$, we have $N_G(Q_1 \cap Q_2) \not\subseteq (Q_1 \cup Q_2)$, contradicting the assumption that both $Q_1$ and $Q_2$ belong to ${\cal S}$. Hence, we have $v \in Q_2$. But then the nodes $u, v \in Q_2$ are distinguished by $Q_1$ and $uv \notin E^+$, a contradiction.

\medskip\noindent
It follows that each node $u \in \bar Q$ belongs to exactly one $S$-articulation clique, say $Q(u)$. Moreover, since each edge $uz$ with $z \in Q \setminus \bar Q$ belongs to $F$ we have $Q \setminus \bar Q \subseteq Q(u)$. Since no $S$-articulation clique contains $\bar Q$ and each node in $\bar Q$ belongs to some $S$-articulation clique, we have that there exist at least two different $S$-articulation cliques $Q_1$, $Q_2$ containing nodes of $\bar Q$. Moreover, since $Q \setminus \bar Q$ is contained both in $Q_1$ and in $Q_2$, we have $\bar Q \subseteq Q_1 \cup Q_2$ for, otherwise, $N_G(Q_1 \cap Q_2) \not\subseteq (Q_1 \cup Q_2)$, contradicting the assumption that both $Q_1$ and $Q_2$ belong to ${\cal S}$. Since no $S$-articulation clique contains $\{x_1, x_2, x_3\}$ we can assume, without loss of generality, $Q(x_1) = Q(x_2) = Q_1$ and $Q(x_3) = Q_2$. If $y_1$ belongs to $L$ then there exists a $S$-articulation clique $Q_0 \in {\cal S}$ containing $x_1$ and not containing $x_2 \notin N^+(y_1)$. But then $Q_0$ distinguishes $x_1$ and $x_2$ in $Q_1$ and $x_1 x_2 \notin E^+$, a contradiction. It follows that $y_1$ does not belong to $L$ and the edge $x_1 y_1$ belongs to $E_{\cal S}$. Consequently, by Lemma~\ref{ThreeEdges}, the edges $x_2 y_1$ and $x_3 y_1$ do not belong to $E$. It follows that $y_1 \notin Q_1$ and the nodes $y_1, x_3 \in N_G(Q_1)$ are non-adjacent and have the common neighbor $x_1 \in Q_1$ in $G$, contradicting the assumption that $Q_1$ is weakly normal in $G$. The lemma follows.
\done

\begin{Lemma}\label{UniqueSArticulation}
Let $Q$ be a $S$-articulation clique in $G$ and let $K \in {\cal K}$ be a spike of $Q$. If there exists a node $y$ which is neither complete nor anti-complete to $K$ in $G^+$ then any edge $uv \in E$ with $u \in K$ and $v \notin Q$ belongs to $E^+$. 
\end{Lemma}

\noindent
\emph{Proof.} Assume, by contradiction, that there exists an edge $uv \in E \setminus E^+$ with $u \in K$ and $v \notin Q$. It follows that there exists a $S$-articulation clique $\bar Q \ne Q$ such that $u, v \in \bar Q$ and hence $K \cap \bar Q \ne \emptyset$. By Lemma~\ref{SpikeInQ} we have $K \subseteq \bar Q$ and hence $K \subseteq Q \cap \bar Q$. Since $y$ is not complete nor anti-complete to $K$, we have $y \notin Q \cup \bar Q$ and $y \in N_G(Q \cap \bar Q)$. This contradicts the assumption that both $Q$ and $\bar Q$ are $S$-articulation cliques. The lemma follows.
\done

\begin{Lemma}\label{RigidPath}
Let $Q$ be a $S$-articulation clique in $G$ and let $K \in {\cal K}$ be a spike of $Q$ with the property that every net triangle $T$ in $G^+$ with $T \cap K \ne \emptyset$ is contained in $K$. Let $\bar U_1$ and $\bar U_2$ be disjoint sets in $N_G(Q)$ with $\bar U_1$ anti-complete to $\bar U_2$ in $G^+$. If $U_1 \equiv N^+(\bar U_1) \cap K$ and $U_2 \equiv N^+(\bar U_2) \cap K$ are both non-empty and disjoint then there exists a rigid edge with one end-node in $U_1$ and the other in $U_2$.
\end{Lemma}

\begin{figure}[htbp]
\centering
\begin{tikzpicture}
\GraphInit[vstyle=Normal]

\tikzstyle{VertexStyle}=[shape = circle, fill = white, minimum size = 20pt, text = black, draw]
\Vertex[L=$w_0$, x=0, y=3]{0}
\Vertex[L=$w_1$, x=2, y=3]{1}
\Vertex[L=$w_2$, x=4, y=3]{2}
\Vertex[L=$w_3$, x=6, y=3]{3}
\Vertex[L=$w_4$, x=8, y=3]{4}
\Vertex[L=$x_0$, x=1, y=0]{5}
\Vertex[L=$x_1$, x=3, y=1]{6}
\Vertex[L=$x_2$, x=5, y=1]{7}
\Vertex[L=$x_3$, x=7, y=0]{8}

\node at (0.25,-0.25) {$U_1$};
\node at (7.75,-0.25) {$U_2$};
\node at (4,-0.5) {$K$};

\draw [blue] plot [smooth cycle] coordinates {(1,-1) (0,0) (1,1) (2,0)}; 
\draw [blue] plot [smooth cycle] coordinates {(7,-1) (6,0) (7,1) (8,0)}; 
\draw [red] plot [smooth cycle] coordinates {(1,-1.5) (-0.5,0) (1,2) (7,2) (8.5,0) (7,-1.5)}; 

\Edge(0)(1) \Edge(0)(5)
\Edge(1)(2) \Edge(1)(5) \Edge(1)(6)
\Edge(2)(3) \Edge(2)(6) \Edge(2)(7)
\Edge(3)(4) \Edge(3)(7) \Edge(3)(8)
\Edge(4)(8)
\Edge(5)(7) \Edge(5)(8)
\Edge(6)(8)

\SetUpEdge[lw = 2pt, color = red]
\Edge(5)(6) \Edge(6)(7) \Edge(7)(8)

\end{tikzpicture}
\captionsetup{justification=centering, margin=1.5cm}
\caption{Proof of Lemma~\ref{RigidPath} (case $p = 3$), in red the rigid edges}
\label{FigRigidPath}
\end{figure}
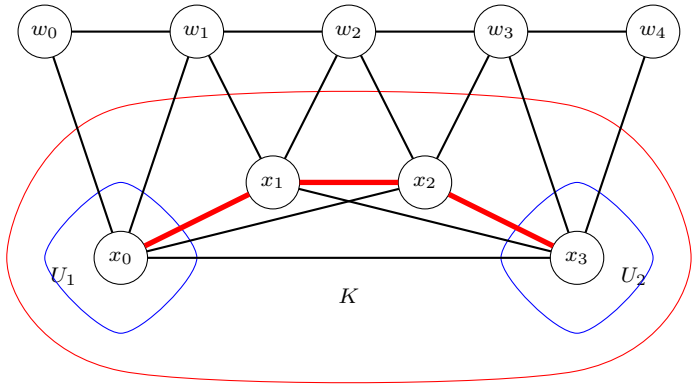

\noindent
\emph{Proof.} Let $q_K$ be the lifting node corresponding to $K$. Observe that, since $K$ is a spike in $G$, the graph $G_R[K]$ is connected. Let $P = (x_0, x_1, \dots, x_p)$ be the shortest path in $G_R[K]$ connecting a node in $U_1$ to a node in $U_2$ (see figure~\ref{FigRigidPath}). By minimality of $P$, each node $x_h \in P$ with $h \in \{1, \dots, p-1\}$ belongs to $K \setminus (U_1 \cup U_2)$. By hypothesis there exist nodes $w_0 \in \bar U_1 \cap N^+(x_0)$ and $w_{p+1} \in \bar U_2 \cap N^+(x_p)$ with $w_0$ and $w_{p+1}$ non-adjacent in $G^+$. For each $h \in \{1, \dots, p\}$ there exists some node $w_h \notin Q$ adjacent to $x_{h-1}$ and $x_h$ in $G$ ($x_{h-1} x_h$ is rigid in $G$). Since $w_0 \in \bar U_1$ is adjacent to $U_1 \subseteq K$ and anti-complete to $U_2 \subseteq K$ we have that $w_0$ is neither complete nor anti-complete to $K$ in $G^+$ and, by Lemma~\ref{UniqueSArticulation}, each edge in $E$ with one end-node in $K$ and the other not in $Q$ belongs to $E^+$. It follows that $w_h x_h, w_h x_{h-1} \in E^+$ for $h \in \{1, \dots, p\}$. 

\smallskip\noindent
Since $P$ is a shortest path in $G_R[K]$, we have that any edge $x_{h-1} x_{h+l}$ with $l \ge 1$ is soft in $G$ and hence $N_G(w_h) \cap P = N^+(w_h) \cap P = \{x_{h-1}, x_h\}$. It follows that we have $w_h \ne w_k$, for $h \ne k$. Moreover, we have $w_0 \ne w_1$ (since $w_1$ is adjacent to $x_1 \notin U_1$ in $G^+$, while $N^+(w_0) \cap K \subseteq U_1$). Analogously, $w_{p+1} \ne w_p$. Since $Q$ is weakly normal in $G$, the nodes $w_h, w_{h+1} \in N_G(Q)$ ($h \in \{1, \dots, p-1\}$) with the common neighbor $x_h \in K$ are adjacent in $G$. Moreover, since $w_h x_h, w_{h+1} x_h \in E^+$ we have, by Lemma~\ref{ThreeEdges}, $w_h w_{h+1} \in E^+$. In addition, $w_0, w_1$ with common neighbor $x_0$ and $w_{p+1}, w_p$ with common neighbor $x_p$ are also adjacent in $G^+$. We claim that $\bar P = (w_0, w_1, \dots, w_p, w_{p+1})$ is an induced path in $G^+$. In fact, let $w_h w_k \in E^+$ ($0 \le h < k \le p+1$) be the edge which maximizes $k-h$ and assume, by contradiction, $k-h > 1$. Since $w_0 w_{p+1} \notin E^+$, we have that either $h > 0$ or $k < p+1$. By symmetry we assume, without loss of generality, $h > 0$. The node $w_h$ is adjacent in $G^+$ to $x_h$, $w_{h-1}$ and $w_k$. By definition of $h, k$, we have $w_{h-1} w_k \notin E^+$. Moreover, $w_{h-1} x_h \notin E^+$ and $w_k x_h \notin E^+$ (since $k-h > 1$). But then $(w_h: x_h, w_{h-1}, w_k)$ is a claw in $G^+$, contradicting Lemma~\ref{LiftingClawFree}. It follows that $\bar P$ is an induced path in $G^+$. Moreover, we have $p = 1$ since, otherwise, $(w_1, w_2, x_1: w_0, w_3, q_K)$ would be a net in $G^+$ and the net triangle $T = \{w_1, w_2, x_1\}$ with $T \cap K \ne \emptyset$ and $T \not\subseteq K$ would provide a contradiction. But then the rigid edge $x_0 x_1$ has one end-node in $U_1$ and the other in $U_2$ and the lemma follows.
\done

\medskip

\medskip\noindent
{\bf Proof of Lemma~\ref{LiftingNetFree}}

\medskip\noindent
Let $C$ be a regular connected component of $G^+$ and assume, by contradiction, that $C$ contains some net. By Lemma~\ref{PreservingSArticulation} there exists a spike $K$ of a $S$-articulation clique $Q$ in $G$ containing a net triangle in $C$. 

\medskip\noindent
\emph{Claim (i). Any edge $uv \in E$ with $u \in K$ and $v \notin Q$ also belongs to $E^+$.}

\smallskip\noindent
\emph{Proof.} Since $K$ contains a net triangle, there exists a node $y$ which is neither complete nor anti-complete to $K$ in $G^+$ and hence, by Lemma~\ref{UniqueSArticulation}, the claim follows.

\noindent
\emph{End of Claim (i).}

\medskip\noindent
Let $\bar K = Q \setminus K$ and let $K^+ = K \cup \{q_K\}$ be the clique obtained from $K$ by adding the associated lifting node $q_K$. The clique $K^+$ is a maximal clique in $C$ containing a net triangle. Hence, by \emph{(i)} of Lemma~\ref{NormalNetTriangle}, $K^+$ is normal in $G^+$ and, since $C$ is regular, by \emph{(ii)} of Lemma~\ref{NormalNetTriangle} $K^+$ is weakly normal in $G^+$.

\medskip\noindent
\emph{Claim (ii). A net triangle $T$ in $G^+$ with $T \cap K \ne \emptyset$ is contained in $K$.}

\smallskip\noindent
\emph{Proof.} Suppose, conversely, that there exists a net triangle $T$ in $G^+$ and nodes $u \in T \cap K$ and $v \in T \setminus K$. But then, by Lemma~\ref{PreservingSArticulation}, $T$ is contained in some $S$-articulation clique $\bar Q \in {\cal S}$ different from $Q$. We have that the nodes $u, v \in \bar Q$ are distinguished by $Q$ and the edge $uv$ contradicts Claim~\emph{(i)}. The claim follows.

\noindent
\emph{End of Claim (ii).}

\medskip
\begin{figure}[htbp]
\centering
\begin{tikzpicture}
\GraphInit[vstyle=Normal]

\tikzstyle{VertexStyle}=[shape = circle, fill = white, minimum size = 20pt, text = black, draw]
\Vertex[L=$u_1$, x=0, y=3]{0}
\Vertex[L=$w_{12}$, x=2, y=3]{1}
\Vertex[L=$u_2$, x=4, y=3]{2}
\Vertex[L=$w_{23}$, x=6, y=3]{3}
\Vertex[L=$u_3$, x=8, y=3]{4}
\Vertex[L=$q_1$, x=1, y=0]{5}
\Vertex[L=$q_2$, x=3, y=1]{6}
\Vertex[L=$k_2$, x=5, y=1]{7}
\Vertex[L=$q_3$, x=7, y=0]{8}

\node at (0.25,-0.25) {$U_1$};
\node at (7.75,-0.25) {$U_3$};
\node at (4,-0.5) {$U_2$};
\node at (4,-1.5) {$K$};

\draw [blue] plot [smooth cycle] coordinates {(1,-1) (0,0) (1,1) (2,0)}; 
\draw [blue] plot [smooth cycle] coordinates {(7,-1) (6,0) (7,1) (8,0)}; 
\draw [blue] plot [smooth cycle] coordinates {(3,0) (2,1) (3,2) (5,2) (6,1) (5,0)}; 
\draw [red] plot [smooth cycle] coordinates {(1,-1.5) (-0.5,0) (1,2) (7,2) (8.5,0) (7,-1.5)}; 

\Edge(0)(1) \Edge(0)(5)
\Edge(1)(2) \Edge(1)(5) \Edge(1)(6)
\Edge(2)(3) \Edge(2)(6) \Edge(2)(7)
\Edge(3)(4) \Edge(3)(7) \Edge(3)(8)
\Edge(4)(8)
\Edge(5)(7) \Edge(5)(8)
\Edge(6)(7) \Edge(6)(8)

\SetUpEdge[lw = 2pt, color = red]
\Edge(5)(6) \Edge(7)(8)

\end{tikzpicture}
\captionsetup{justification=centering, margin=1.5cm}
\caption{Proof of Lemma~\ref{LiftingNetFree} after Claim~\emph{(ii)}, the edges in red are rigid}
\label{FigClaim2}
\end{figure}
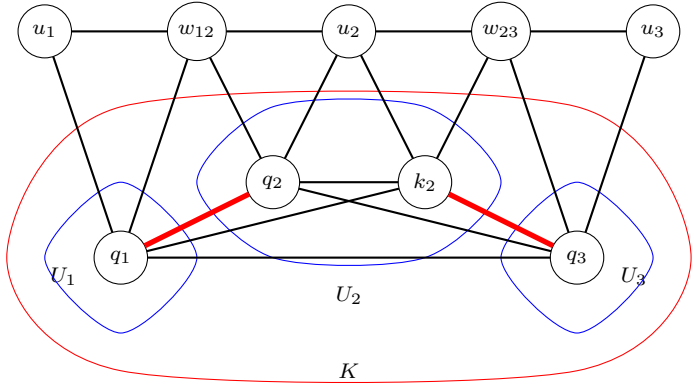

\medskip\noindent
Let $\{u_1, u_2, u_3\}$ be a stable set of $G^+$ in $N^+(K^+)$ (it exists since $K^+$ is normal in $G^+$) and let $U_i = N^+(u_i) \cap K = N_{G}(u_i) \cap K$ ($i = 1, 2, 3$). We have that $U_1, U_2, U_3$ are non-empty and disjoint (since $K^+$ is weakly normal in $G^+$). Letting $\bar U_1 = \{u_1\}$ and $\bar U_2 = \{u_2\}$ we have, by Lemma~\ref{RigidPath}, that there exists in $G$ (and in $G^+$) a rigid edge $q_1 q_2$ with $q_1 \in U_1$ and $q_2 \in U_2$. Analogously, there exists a rigid edge $k_2 q_3$ with $k_2 \in U_2$ and $q_3 \in U_3$ (possibly $k_2 \equiv q_2$). It follows that there exists some node $w_{12} \notin Q$ adjacent to $q_1$ and $q_2$ (both in $G$ and in $G^+$, by Claim~\emph{(i)}). Since $K^+$ is weakly normal in $G^+$, the nodes $u_1, w_{12} \in N^+(K^+)$ with the common neighbor $q_1 \in K^+$ are adjacent in $G^+$. Analogously, the nodes $u_2, w_{12} \in N^+(K^+)$ with the common neighbor $q_2 \in K^+$ are also adjacent. If $w_{12}$ were adjacent to $q_3$ or to $u_3$ in $G^+$ then $(w_{12}: u_1, u_2, q_3)$ or $(w_{12}: u_1, u_2, u_3)$ would be a claw in $G^+$, contradicting Lemma~\ref{LiftingClawFree}. It follows that $w_{12} q_3 \notin E^+$ and $w_{12} u_3 \notin E^+$ (and hence, by Claim~\emph{(i)}, $w_{12} q_3 \notin E$). However, since $k_2 q_3$ is a rigid edge in $G$, there exists some node $w_{23} \notin Q$ adjacent to $k_2$ and $q_3$ (both in $G$ and in $G^+$, by Claim~\emph{(i)}), with $w_{23} \ne w_{12}$. Again, since $K^+$ is weakly normal in $G^+$, the node $w_{23}$ is adjacent to both $u_2$ and $u_3$ in $G^+$. Moreover, $w_{23}$ is non-adjacent to $q_1$ (both in $G$ and in $G^+$, by Claim~\emph{(i)}) and to $u_1$ in $G^+$ (see figure~\ref{FigClaim2}). Assume that $w_{12} w_{23} \notin E^+$. Hence $w_{23} q_2 \notin E^+$ for, otherwise, $(q_2: q_1, w_{12}, u_2, w_{23}, q_3)$ would be a $5$-wheel in $C$, contradicting the assumption that $C$ is a regular connected component of $G^+$. But then $(w_{12}, q_2, u_2: u_1, q_K, w_{23})$ is a net in $G^+$ and $T = \{w_{12}, q_2, u_2\}$ is a net triangle in $G^+$ with $T \cap K \ne \emptyset$ and $T \not\subseteq K$, contradicting Claim~\emph{(ii)}. 

\smallskip\noindent
Hence, we have $w_{12} w_{23} \in E^+$. It follows that $w_{23} q_2 \in E^+$ for, otherwise, $(w_{12}: u_1, w_{23}, q_2)$ would be a claw in $G^+$. But then $(w_{12}, q_2, w_{23}: u_1, q_K, u_3)$ is a net in $G^+$ and $T = \{w_{12}, q_2, w_{23}\}$ is a net triangle in $G^+$ with $T \cap K \ne \emptyset$ and $T \not\subseteq K$, again contradicting Claim~\emph{(ii)}. The lemma follows.
\done

\medskip\noindent
In \cite{FaenzaOS14} Faenza, Oriolo and Stauffer introduce the concept of \emph{hyper-line strip}. A hyper-line strip $(H,{\cal A})$ is characterized by an induced subgraph $H$ of $G$ and a family ${\cal A}$ of disjoint non-empty cliques contained in $V(H)$ with $1 \le |{\cal A}| \le 2$ (\emph{extremities}). For each extremity $A \in {\cal A}$ the set $A \cup (N(A) \setminus V(H))$ is an articulation clique. Moreover, the \emph{core} $C(H, {\cal A})$ of the hyper-line strip, consisting of the nodes in $V(H)$ that do not belong to the extremities, is anticomplete to $V \setminus V(H)$. Faenza, Oriolo and Stauffer show that each irregular node of a claw-free graph $G(V, E)$ with $\alpha(G) \ge 4$ (hub of a $5$-wheel by \cite{Fouquet} Corollary~1.2) is contained in a hyper-line strip. More precisely, the arguments used in the proof of Lemma~6.4 (\cite{FaenzaOS14}) and in particular the proof of Claim~20 imply the following theorem: 

\begin{Theorem}\label{FaenzaEtAl}
Let $G(V, E)$ be a connected claw-free graph with $\alpha(G) \ge 4$. Then there exists a family ${\cal H}$ of node-disjoint hyper-line strips with the property that every $5$-wheel $W = (w: v_1, \dots, v_5)$ of $G$ is contained in some $H \in {\cal H}$ with $\alpha(H) \le 3$. Moreover, $V(H) \subseteq N[W]$, and each node $z \in V(H)$ is not simplicial.
\done
\end{Theorem}

\medskip
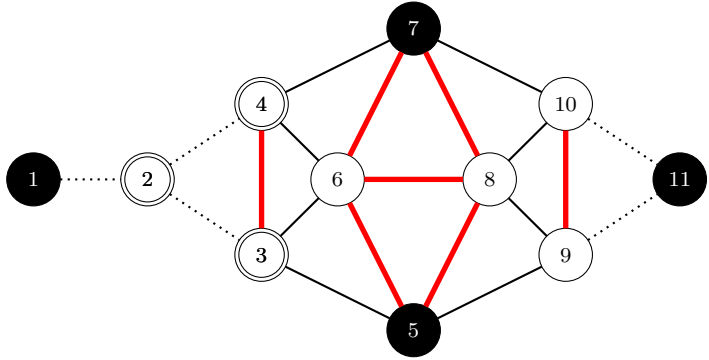
\begin{figure}[htbp]
\centering
\begin{tikzpicture}
\GraphInit[vstyle=Normal]

\tikzstyle{VertexStyle}=[shape = circle, fill = black, minimum size = 20pt, text = white, draw]
\Vertex[L=$1$, x=0, y=2]{1}
\Vertex[L=$5$, x=5, y=0]{5}
\Vertex[L=$7$, x=5, y=4]{7}
\Vertex[L=$11$, x=8.5, y=2]{11}

\tikzstyle{VertexStyle}=[shape = circle, fill = white, minimum size = 20pt, text = black, draw]
\Vertex[L=$2$, x=1.5, y=2]{2}
\Vertex[L=$3$, x=3, y=1]{3}
\Vertex[L=$4$, x=3, y=3]{4}
\Vertex[L=$6$, x=4, y=2]{6}
\Vertex[L=$8$, x=6, y=2]{8}
\Vertex[L=$9$, x=7, y=1]{9}
\Vertex[L=$10$, x=7, y=3]{10}

\tikzstyle{VertexStyle}=[shape = circle, minimum size = 17pt, draw]
\Vertex[L=$2$, x=1.5, y=2]{}
\Vertex[L=$3$, x=3, y=1]{}
\Vertex[L=$4$, x=3, y=3]{}

\Edge(3)(5) \Edge(3)(6)
\Edge(4)(6) \Edge(4)(7)
\Edge(5)(9)
\Edge(7)(10)
\Edge(8)(9) \Edge(8)(10)

\SetUpEdge[style=dotted]
\Edge(1)(2)
\Edge(2)(3) \Edge(2)(4)
\Edge(9)(11)
\Edge(10)(11)

\SetUpEdge[lw = 2pt, color = red]
\Edge(3)(4)
\Edge(5)(6) \Edge(5)(8)
\Edge(6)(7) \Edge(6)(8)
\Edge(7)(8)
\Edge(9)(10)

\end{tikzpicture}
\captionsetup{justification=centering, margin=0.5cm}
\caption{Rigid structure and ungluing of graph containing a hyperline strip}
\label{FigIrregular}
\end{figure}

\noindent
\textsl{
\small
In the graph shown in figure~\ref{FigIrregular}, $S = \{1, 5, 7, 11\}$. The cliques in ${\cal C} \cup {\cal F}$ are $C_1 = \{1, 2\}$, $C_5 = \{3, 5, 6\}$, $\bar C_5 = \{5, 8, 9\}$, $C_7 = \{4, 6, 7\}$, $\bar C_7 = \{7, 8, 10\}$, $C_{11} = \{9, 10, 11\}$, $F = \{2, 3, 4\}$. The family ${\cal S}$ contains the cliques $C_1$, $C_{11}$ and $F$. The edges marked in red are rigid while the edges in solid black are soft but are not contained in any clique of ${\cal S}$ and hence belong to $E_{\cal S}$. The dotted edges belong to $E \setminus E_{\cal S}$ since they are soft, belong to some clique $K$ of ${\cal S}$ and connect different components of $G_R[K]$. Finally, the subgraph induced by $H = \{3, 4, 5, 6, 7, 8, 9, 10\}$ is a hyperline strip $(H, {\cal A})$ containing the irregular nodes $6$ and $8$ and the corresponding family ${\cal A}$ is composed by $A_1 = \{3, 4\}$ and $A_2 = \{9, 10\}$.
}

\medskip\noindent
Building upon this crucial result we can prove the following:

\begin{Lemma}\label{IrregularComponent}
If $W = (w: v_1, \dots, v_5)$ is a $5$-wheel in $G$ then there exists a strip $C$ of $G_{\cal S}$ with the property that $W \subseteq V(C)$ and $\alpha(C) \le 3$. 
\end{Lemma}

\noindent
\emph{Proof.} By Theorem~\ref{FaenzaEtAl} we have that there exists a hyper-line strip $(H, {\cal A})$ with $W \subseteq V(H)$ and $\alpha(H) \le 3$. Since $G$ is connected and $\alpha(G) \ge 4$, at least one of the extremities, say $A_1 \in {\cal A}$ has $N_G(A_1) \setminus V(H) \ne \emptyset$. Let $A_2$ be the other extremity of $H$ (if it exists) and denote by $K_i$ the articulation clique (and hence $S$-articulation by Lemma~\ref{ArticulationSArticulation}) $A_i \cup (N_G(A_i) \setminus V(H))$ ($i = 1, 2$). Denote by $R = \{v_1, \dots, v_5\}$ the rim of $W$.

\smallskip\noindent
Observe first that the node $w$ does not belong to a $S$-articulation clique (each clique containing $w$ is not weakly normal) and hence the edges $w v_h$ belong to $E^+$ for each $v_h \in R$. Moreover, each edge $v_h v_{h+1}$ (sums taken modulo $5$) also belongs to $E^+$, otherwise $(w: v_h, v_{h+1}, v_{h+3})$ would be a claw in $G^+$, contradicting Lemma~\ref{LiftingClawFree}. It follows that the $5$-wheel $W$ is an induced subgraph of $G^+$.

\smallskip\noindent
Let $uv \in E$ be an edge with $u \in V(H)$ and $v \notin V(H)$. We have $u \in A_i$ and $v \in N_G(A_i) \setminus V(H)$ (for some $i \in \{1, 2\}$). Without loss of generality, assume $u \in A_1$. Suppose that $uv$ is rigid. It follows that there exists a node $z \in N_G(K_1)$ adjacent to both $u$ and $v$. Since the node $z$ does not belong to $K_1$, it belongs to $V(H) \setminus A_1$. Moreover, since the core of $H$ is anticomplete to $V \setminus V(H)$, we have $z \in A_2$. But then $v \in N_G(A_2) \setminus V(H)$ belongs to $K_2$ and $K_2$ distinguishes $u$ and $v$, a contradiction. Hence each edge $uv$ with $u \in V(H)$ and $v \notin V(H)$ is a soft edge in $G$. It follows that the connected component $C$ of $G_{\cal S}$ containing $W$ is a subgraph of $H$.

\smallskip\noindent
We claim that $C$ contains at most two spikes. Suppose, conversely that there exists three spikes in $C$ and let $y_1, y_2, y_3$ be the corresponding lifting nodes in $G^+$. Since $C$ is a subgraph of $H$ we have $C \subseteq N_G(W)$ and hence $y_1, y_2, y_3$ are at most at distance three from $w$. Moreover, since $w$ is not contained in any $S$-articulation clique, $y_1, y_2, y_3 \notin N^+(w)$. If $y_i \in N^+(R)$, we have that $y_i$ is adjacent to exactly two consecutive nodes in $R$. If, on the other hand, $y_i \notin N^+(R)$, we have that $y_i$ is adjacent to some node $u_i$ which is adjacent to exactly two consecutive nodes in $R$ and non-adjacent to $w$. 

\smallskip\noindent
If $y_1, y_2, y_3$ are adjacent to $R$ in $G^+$ then two of them have a common neighbor in $R$, contradicting Lemma~\ref{SpikesAreDisjoint}. If two of the lifting nodes, say $y_1, y_2$, belong to $N^+(R)$, we have that one of them (say $y_1$) and $u_3$ have a common neighbor in $R$, say $v_h$. But then $(v_h: y_1, u_3, w)$ is a claw in $G^+$, a contradiction. If only one lifting node, say $y_1$ is adjacent to $R$, we have that either $y_1, u_3$ or $y_1, u_2$ or $u_2, u_3$ have a common neighbor in $R$. In the first two cases, as above, there is a claw in $G^+$, a contradiction. If $u_2, u_3$ have a common neighbor in $R$, say $v_h$, assume, without loss of generality, $N^+(u_2) \cap R = \{v_{h-1}, v_h\}$ and $N^+(u_3) \cap R = \{v_h, v_{h+1}\}$. But then $u_2, u_3$ are adjacent in $G^+$ (otherwise $(v_h: u_2, u_3, w)$ is a claw in $G^+$) and, consequently, $(u_2: v_{h-1}, y_1, u_3)$ is a claw in $G^+$, a contradiction.  Finally, if $y_1, y_2, y_3$ are non-adjacent to $R$ in $G^+$ then two of the nodes $u_1, u_2, u_3$ have a common neighbor in $R$ and, as above, there is a claw in $G^+$, a contradiction. This implies that $C$ is a connected component of $G_{\cal S}$ containing at most two spikes and hence it is a strip. The lemma follows.
\done

\bigskip\noindent
{\bf Proof of Theorem~\ref{S-StripTheorem}}

Let $C$ be any connected component of $G_{\cal S}$. Since $G$ is a connected claw-free graph with $\alpha(G) \ge 4$, by \cite{Fouquet} (Corollary~1.2) every irregular node of $G$ is the hub of a $5$-wheel. Hence, if $C$ is an irregular connected component the theorem follows by Lemma~\ref{IrregularComponent}. Hence suppose that $C$ is a regular connected component of $G_{\cal S}$. Let $C^+$ be the connected component of $G^+$ obtained by adding to $C$ the associated lifting nodes, $S^+$ the extension of $S$ in $G^+$ and let $S^+_C = C^+ \cap S^+$. By Lemma~\ref{LiftingClawFree} and Lemma~\ref{LiftingNetFree}, $G^+[C^+]$ is a \{claw, net\}-free graph. If $|S^+_C| \ge 4$ we have (\cite{NobiliSassano16a} (Theorem~2.1)) that each node $s \in S^+_C$ defines at most two wings in $G^+$ with respect to $S^+_C$. If, on the other hand, $|S^+_C| \le 3$ then trivially each node $s \in S^+_C$ defines at most two distinct wings. Hence, we can conclude that any node $s \in S^+_C$ defines at most two wings in $G^+$. Let $H(S^+_C, T)$ be the graph where $xy$ is an edge in $T$ if and only if the nodes $x, y \in S^+_C$ are the extrema of a wing $W(x, y)$ in $G^+$. Observe that each node $s \in S^+_C$ has degree either $0$, $1$ or $2$ in $H$.

\medskip\noindent
\emph{Claim (i). The graph $H$ is connected.}

\smallskip\noindent
\emph{Proof.} Assume conversely that there exist at least two connected components of $H$. Since $C^+$ is a connected component in $G^+$, each pair of nodes in $S^+_C$ is connected by a (shortest) path in $G^+$. Let $P = (x, z_1, \dots, z_h, y)$ ($h \ge 1$) be a shortest path connecting two different components of $H$. Let $C_1$ and $C_2$ be the components of $H$ connected by $P$, with $x \in C_1$ and $y \in C_2$. Observe that, since the nodes in $L$ are simplicial in $G^+$, the internal nodes of $P$ belong to $V$. Moreover, by minimality of $P$, $z_i \notin S^+_C$ ($i = 1, \dots, h$) and $h \ge 2$, otherwise $W(x, y)$ would be a wing, contradicting the assumption that $x$ and $y$ do not belong to the same connected component of $H$. If $h \ge 3$, since $z_2$ is not superfree with respect to $S^+$ we have that there exists at least one node $t \in S^+$ adjacent to $z_2$ and different from $x$ and $y$; we have $t \in S^+_C$. If $t \notin C_1$ then $t z_1 \notin E^+$ and the path $(x, z_1, z_2, t)$ would contradict the minimality of $P$. It follows that $t$ belongs to $C_1$ and hence $t, y$ are in different connected components of $H$. But then $(t, z_2, z_3, \dots, z_h, y)$ is a path connecting two nodes in different components of $H$ which is shorter than $P$, a contradiction. Consequently, we have $h = 2$. If $z_1$ is a bound node then there exists a node $t \in S^+$ adjacent to $z_1$ and different from $x$ and $y$. The node $t$ belongs to $S^+_C$ and, by claw-freeness, is also adjacent to $z_2$. It follows that both $W(x, t)$ and $W(y, t)$ are wings and hence $t$ belongs to both $C_1$ and $C_2$, a contradiction. Hence $z_1$ and, by symmetry, $z_2$ are free nodes. But then $x$ and $y$ define a wing, again a contradiction.

\noindent
\emph{End of Claim (i).}

\medskip\noindent
By what proved above $H$ is either an isolated node or a path with at least one edge or a cycle. Observe that for each node $u \in L$ either $u$  belongs to $S^+_C$ or the spike complete to $u$ contains a node in $S^+_C$, so $|C^+ \cap L| \le |S^+_C|$. Hence, if $H$ is a singleton then $C$ is a strip. Consequently, we can assume $|S^+_C| = p \ge 2$. In this case there exists an ordering $\{s_1, s_2, \dots, s_p\}$ of $S^+_C$ such that each $s_i$ ($2 \le i \le p-1$) defines wings with $s_{i-1}$ and with $s_{i+1}$. If $H$ is a cycle, then also $W(s_1, s_p)$ is a wing and every node in $S^+_C$ defines wings with exactly two other distinct nodes in $S^+_C$. 

\medskip\noindent
\emph{Claim (ii). If there exists a node $u \in C^+ \cap L$ such that the unique node $\bar u \in S^+_C \cap N^+[u]$ defines two wings in $G^+$ then $C^+ \cap L = \{u\}$}

\smallskip\noindent
\emph{Proof.} Let $K = N^+(u) \in {\cal K}$ be the spike complete to $u$ in $G^+$ and $Q \in {\cal S}$ the clique of $G$ containing $K$. Observe that the node $u$ either belongs to $S^+_C$ and coincides with $\bar u$ or is free and adjacent to $\bar u \in K$. Let $W(t_1, \bar u)$ and $W(t_2, \bar u)$ be the wings defined by $\bar u$ in $G^+$. Let $\bar U_i$ be the set containing $t_i$ and the free nodes in $W(t_i, \bar u) \cap N^+(t_i)$ ($i = 1, 2$). Since the nodes in $\bar U_i$ are non-adjacent to $\bar u$ in $G^+$, we have $\bar U_i \cap K = \emptyset$ ($i = 1, 2$). Moreover, $\bar U_1 \cap \bar U_2 = \emptyset$. Let $U_i = N^+(\bar U_i) \cap K$ ($i = 1, 2$). Observe that we have $U_i \equiv W(t_i, \bar u) \cap K$ ($i = 1, 2$) and hence $U_1, U_2$ are non-empty. If there exists a node $x \in U_1 \cap U_2$, then it belongs to $W(t_1, \bar u) \cap W(t_2, \bar u)$ and hence is a free node in $N^+(\bar u)$. In this case, there exist free nodes $x_1 \in N^+(t_1) \cap N^+(x)$ and $x_2 \in N^+(t_2) \cap N^+(x)$. Moreover, we have $x_1 x_2 \in E^+$ (otherwise $(x: \bar u, x_1, x_2)$ is a claw in $G^+$, contradicting Lemma~\ref{LiftingClawFree}). But then $(x_1, x_2, x: t_1, t_2, \bar u)$ is a net in $G^+[C]$, contradicting Lemma~\ref{LiftingNetFree}. It follows that $U_1$ and $U_2$ are disjoint. Moreover $K$ does not intersect any net triangle (again by Lemma~\ref{LiftingNetFree}). Hence, by Lemma~\ref{RigidPath}, we have that there exists in $G$ (and in $G^+$) a rigid edge $v_1 v_2$ with $v_1 \in U_1$ and $v_2 \in U_2$. Moreover, if $v_1$ is bound let $\bar v_1 \equiv t_1$, otherwise let $\bar v_1$ be a free node in $N^+(t_1) \cap N^+(v_1)$. Analogously, define $\bar v_2$. Observe that $\bar v_i$ belongs to $\bar U_i$ ($i = 1, 2$) and hence it is neither complete nor anti-complete to $K$ in $G^+$. Let $w \notin Q$ be a node adjacent both to $v_1$ and to $v_2$ in $G$ (it exists since $v_1 v_2$ is a rigid edge). Since $\bar v_1$ is neither complete nor anti-complete to $K$ in $G^+$, by Lemma~\ref{UniqueSArticulation} we have $w v_1, w v_2 \in E^+$. Moreover, $w \bar v_1 \in E^+$ (otherwise $(v_1: u, \bar v_1, w)$ is a claw in $G^+$, contradicting Lemma~\ref{LiftingClawFree}). Analogously, $w \bar v_2 \in E^+$.

\smallskip\noindent
If $\bar v_1 \ne t_1$ we have that $(v_1, \bar v_1, w: u, t_1, \bar v_2)$ is a net in $G^+[C^+]$, contradicting Lemma~\ref{LiftingNetFree}. It follows that $\bar v_1 \equiv t_1$ and, analogously, $\bar v_2 \equiv t_2$. Hence we have that $W(t_1, t_2)$ is a wing in $G^+$ with respect to $S^+_C$ (it contains $w$) and each pair in $\{t_1, t_2, \bar u\}$ defines a wing. Consequently $S^+_C = \{t_1, t_2, \bar u\}$ and $H$ is a triangle. Suppose now that $|C^+ \cap L| \ge 2$ and let $y \ne u$ be a node in $C^+ \cap L$. The node $y$ is non-adjacent to $\bar u$. Consequently, since $N^+[y] \cap S^+_C$ is non-empty and $t_1, t_2 \notin L$ (they are adjacent to the spike $K$ and $q_K \equiv u$), we have that $y$ is adjacent either to $t_1$ or $t_2$. Without loss of generality we can assume $y t_1 \in E^+$ and hence $y t_2 \notin E^+$. The node $y$ is non-adjacent to $v_1$ (since $v_1 \in K$) and non-adjacent to $w$ (otherwise $(w: y, v_1, t_2)$ would be a claw in $G^+$). But then $(t_1, v_1, w: y, u, t_2)$ is a net in $G^+[C^+]$, contradicting Lemma~\ref{LiftingNetFree}. Hence, $C^+ \cap L = \{u\}$ and the claim follows.

\noindent
\emph{End of Claim (ii).}

\medskip\noindent
By the above claim, if $|C^+ \cap L| \ge 2$ then the nodes in $C^+ \cap L$ can only belong to $N^+[s_1] \cup N^+[s_p]$. Since $s_1$ ($s_p$) either belongs to $L$ or is adjacent in $G^+$ to at most one node in $L$ we have $|C^+ \cap L| \le 2$ and hence $C$ is a strip. The theorem follows.
\done

\section{Finding a canonical stable set and $S$-articulation cliques in ${\cal O}(|V|^2)$}\label{Finding}

\noindent
In what follows, we assume that the graph $G(V, E)$ is represented by an array of $|V|$ records, each one associated with a node and containing the list of its neighbors (sorted with respect to some given linear ordering on $V$) and the information pertaining to that node. In particular, each node $v \in V$ is labeled as stable, bound or free with respect to a canonical stable set $S$. In addition, the list $S(v)$ (containing at most two elements) is available for each node $v \in V \setminus S$. Observe that the above mentioned data structures can all be constructed in ${\cal O}(|E|)$ time.

\medskip\noindent
The algorithms described in this section exploits a suitable list ${\cal B}$ of ${\cal O}(|V|)$ cliques in $G$. Such a list contains the cliques in the $S$-cover ${\cal C}$, the F-cliques in ${\cal F}$, the free similarity classes in $\{F(s): s \in S\}$ and a subset of the non-empty intersections of pairs of such cliques. With each clique $Q \in {\cal B}$ is uniquely associated an identifier $id[Q]$ and pertaining information (with size ${\cal O}(1)$) like, for example, whether $Q$ is the intersection of two different cliques (and which are the identifiers of such cliques), whether $Q$ is an F-clique (and, if trivial, which are the identifiers of the two similarity classes it intersects), whether $Q$ is a free similarity class $F(s)$ (and which is the stable node $s$) or whether $Q$ belongs to ${\cal C}$ (and which is the unique stable node in $Q$). The cliques in ${\cal B}$ are accessible through their identifiers and other special collection of cliques are represented as lists of identifiers of cliques in ${\cal B}$ and called \emph{families}. Moreover, for each $s \in S$, the identifier of $F(s)$ is added to the information pertaining to the node $s$.

\medskip\noindent
We start by assessing the complexity of constructing a canonical stable set $S$, the $S$-cover ${\cal C}$, the list of F-cliques ${\cal F}$ and the list of cliques $\{F(s): s \in S\}$.

\begin{Theorem}\label{CanonicalConstruction}
A canonical stable set $S$ of a claw-free graph $G(V, E)$ can be obtained in time ${\cal O}(|E|)$.
\end{Theorem}

\noindent
\emph{Proof}. Let $T$ be any maximal stable set of $G$. Let $P = (x, s, y)$ be an induced $P_3$ in $G$ with $s \in T$ and $x, y \in F(s)$. As customary we say that $P$ is augmenting with respect to $T$ and call the set $\bar T = T \setminus \{s\} \cup \{x, y\}$, which is a stable set, the augmentation of $T$ with respect to $P$.

\medskip\noindent
\emph{Claim (i). $\bar T$ is a maximal stable set, the set of free nodes with respect to $\bar T$ is strictly contained in the set of free nodes with respect to $T$ and every $P_3$ which is augmenting with respect to $\bar T$ is also augmenting with respect to $T$.}

\smallskip\noindent
\emph{Proof.} Suppose first that $\bar T$ is not maximal and hence there exists some node $v \in V \setminus \bar T$ which is non-adjacent to every node in $\bar T$. In particular, $v \notin N(x) \cup N(y)$. Moreover, since $T$ is maximal, we have $v \in N(s)$. But then $(s: x, y, v)$ is a claw in $G$, a contradiction. Suppose now that there exists some node $z$ which is free with respect to $\bar T$ but is not free with respect to $T$. Since $s$ is bound with respect to $\bar T$ and any other node in $T$ also belongs to $\bar T$ we have $z \notin T$; moreover, since $T$ is maximal, $z$ is not superfree and hence is bound with respect to $T$. The node $z$ is adjacent to $s$ (otherwise it would be bound also with respect to $\bar T$) and to some other stable node $\bar s \in T \cap \bar T$. Moreover, since $z$ is free with respect to $\bar T$ and is adjacent to $\bar s \in \bar T$, it is non-adjacent to $x$ and to $y$. But then $(s: x, y, z)$ is a claw in $G$, a contradiction. Suppose now that there exists a $P_3$ $(p, t, q)$ which is augmenting with respect to $\bar T$ but not with respect to $T$. Since the set of free nodes with respect to $\bar T$ is contained in the set of free nodes with respect to $T$, we have that $p$ and $q$ are also free with respect to $T$. Hence, we have $t \in \bar T \setminus T \equiv \{x, y\}$ (otherwise $(p, t, q)$ would be augmenting with respect to $T$). Moreover, $p$ and $q$ are non-adjacent to any node in $T \cap \bar T$ (otherwise they would not be free with respect to $\bar T$) and hence are both adjacent to $s$ (otherwise they would not be free with respect to $T$). Hence, without loss of generality, we can assume $t \equiv x$ and so $y$ is non-adjacent to both $p$ and $q$ (otherwise they would not be free with respect to $\bar T$). But then $(s: p, q, y)$ is a claw in $G$, a contradiction.

\noindent
\emph{End of Claim (i).}

\medskip\noindent
Let $S_0 = \{s_1, s_2, \dots, s_q\}$ be a maximal stable set of $G(V, E)$ and $F(S_0)$ the associated set of free nodes. Observe that the data structures representing the sets $S_0$ and $F(s)$ for all $s \in S_0$ can be constructed in ${\cal O}(|E|)$ time. We now prove that a canonical stable set $S_q$ can be obtained from $S_0$ by iteratively looking for a possible augmentation of a current stable set $S_{i-1}$ and producing the updated stable set $S_i$ ($i$ initially set to $1$), in overall time ${\cal O}(|E|)$. At stage $i$ of the procedure we examine the node $s_i \in S_0$. Let $G_i(V_i, E_i)$ be the subgraph of $G$ induced by $N[s_i]$. Observe that, by Lemma~4 in \cite{KKM00}, $|V_i| = {\cal O}(\sqrt{|E_i|})$. We scan the set $V_i \cap F(S_{i-1})$ looking for a pair of non-adjacent nodes. This can be done in time ${\cal O}(|E_i|)$. If we find an augmenting $P_3$ $(x_i, s_i, y_i)$, we update the stable set $S_{i-1}$ by letting $S_i := S_{i-1} \setminus \{s_i\} \cup \{x_i, y_i\}$, otherwise we set $S_i := S_{i-1}$. Moreover, the set $F(S_i)$ of the free nodes with respect to $S_i$ is obtained by updating $F(S_{i-1})$. In particular, if $S_i \equiv S_{i-1}$ trivially we have $F(S_i) \equiv F(S_{i-1})$, otherwise, by Claim~\emph{(i)}, we only have to remove from $F(S_{i-1})$ the nodes which are not free with respect to $S_i$. Observe that the nodes to be removed are $x_i$, $y_i$ (which become stable) and any node which becomes bound (either adjacent to both $x_i$ and $y_i$ or adjacent to $x_i$ or $y_i$ and not to $s_i$). It follows that $F(S_i)$ can be obtained from $F(S_{i-1})$ by checking the nodes in the neighborhood of $x_i$ and $y_i$ in time ${\cal O}(|N(x_i) \cup N(y_i)|)$. Observe that Claim~\emph{(i)} ensures that no new augmenting $P_3$ is produced by the operation. This implies that we have only to check the nodes in $S_0$ as stable nodes in augmenting $P_3$. Let $A \subseteq \{1, 2, \dots, q\}$ be the set of indices of the iterations that produced an augmentation. It follows that the overall complexity of the procedure is ${\cal O}(\sum_{i = 1}^{q}|E_i| + \sum_{i \in A}(|N(x_i) \cup N(y_i)|)$. By claw-freeness, each edge in $E$ belongs to at most two sets $E_i$ and hence $\sum_{i = 1}^{q}|E_i| \le 2|E|$. Moreover, the neighborhood of the nodes $x_i$ and $y_i$, removed from the set of free nodes at stage $i$, will not be scanned again in the subsequent stages and hence $\sum_{i \in A}(|N(x_i) \cup N(y_i)|) \le \sum_{v \in V} |N(v)| = 2|E|$. Consequently, the overall complexity of the procedure is ${\cal O}(|E|)$.
\done

\begin{Theorem}\label{SCoverFreeComponentComplexity}
Let $G(V, E)$ be a claw-free graph and $S$ a canonical stable set of $G$. Then the $S$-cover ${\cal C}$, the list of F-cliques ${\cal F}$ and the list of cliques $\{F(s): s \in S\}$ can be constructed and added to ${\cal B}$ in ${\cal O}(|V|^2)$ time.
\end{Theorem}

\noindent
\emph{Proof}. We first show that the $S$-cover ${\cal C}$ can be constructed in ${\cal O}(|V|^2)$ time. In fact, for each $s \in S$, we try to bi-color the complement $\bar G_s$ of $G[N(s)]$ by performing a breadth first search in ${\cal O}(|N(s)|^2)$ time. The node $s$ is strongly regular if and only if we get a partition $(A, B, I)$ of $N(s)$ ($A, B$ or $I$ possibly empty) where $\bar G_s[A \cup B]$ is connected and bipartite and the nodes in $I$ are isolated in $\bar G_s$. If the partition exists and $A, B$ are non-empty, the maximal cliques $C_s, \bar C_s$ in the unique pair covering $N[s]$ are induced by $A \cup I \cup \{s\}$ and $B \cup I \cup \{s\}$; if all the nodes in $N(s)$ are isolated in $\bar G_s$ then the unique maximal clique covering $N[s]$ is $C_s \equiv \bar C_s = I \cup \{s\}$ (see also \cite{FaenzaOS14} Lemma~3.3). Moreover, since each node $v \in V$ is adjacent to at most two nodes in $S$, we have $\sum_{s \in S} |N(s)|^2 \le 4 |V|^2$ and hence the cliques in ${\cal C}$ can be constructed and added to ${\cal B}$ in ${\cal O}(|V|^2)$ time. As to ${\cal F}$ and $\{F(s): s \in S\}$, observe that in ${\cal O}(|E|)$ time we can construct the set $F(S)$ of free nodes with respect to $S$ and partition $F(S)$ into the free similarity classes $F(s)$ ($s \in S$) that are added to ${\cal B}$. In turn, this partition allows us to construct in ${\cal O}(|E|)$ time the free dissimilarity graph of $G$ and hence, in ${\cal O}(|V|^2)$ time, the list of the connected components of such a graph which are cliques in $G$. Finally, again in ${\cal O}(|V|^2)$ time, we can remove from such a list the cliques which are not maximal in $G$ thus obtaining ${\cal F}$ that is added to ${\cal B}$.
\done

\medskip\noindent
Observe that, while constructing the lists ${\cal C}$, ${\cal F}$ and $\{F(s): s \in S\}$, we can record for each clique $C$ in ${\cal C}$ the unique stable node in $C$ and, for each F-clique, we can easily check whether it is trivial or not and, in the first case, record the two similarity classes it intersects.

\medskip\noindent
We are now ready to describe the algorithm for constructing the family of $S$-articulation cliques ${\cal S}$. To this purpose, we will make use of the following data structures:

\begin{description}
	 \item[-] ${\cal B}$: the list of cliques in ${\cal C}$, ${\cal F}$, $\{F(s): s \in S\}$ and the cliques $Q_i \cap Q_j$ (if non-empty) with either $Q_i, Q_j \in {\cal C} \cup {\cal F}$ or $Q_i \in {\cal F}$ and $Q_j \in \{F(s): s \in S\}$;
	 \item[-] ${\cal B}_u$ ($u \in V$): the family (list of identifiers) of cliques in ${\cal B}$ containing $u$;
   \item[-] ${\cal C}_u$ ($u \in V$): the family of cliques in ${\cal C}$ containing $u$;
   \item[-] ${\cal F}_u$ ($u \in V$): the family of F-cliques in ${\cal F}$ containing $u$;
	 \item[-] ${\cal D}_s$ ($s \in S$): the family of non-trivial F-cliques in ${\cal F}$ intersecting the similarity class $F(s)$;
	 \item[-] ${\cal D}_{st}$ ($s, t \in S$): the family of trivial F-cliques in ${\cal F}$ intersecting the similarity classes $F(s)$ and $F(t)$;
	 \item[-] $\sigma[id[Q_i], id[Q_j]]$ ($Q_i, Q_j \in {\cal C} \cup {\cal F} \cup \{F(s): s \in S\}$): $id[Q_i \cap Q_j]$ if $Q_i \cap Q_j \in {\cal B}$ and $nil$ otherwise;
	 \item[-] $n[u, id[Q]]$ ($u \in V$ and $Q \in {\cal B}$): the number of nodes in $N(u) \cap Q$.
\end{description}

\medskip\noindent
In what follows, to simplify the notation, we will write $Q$ instead of $id[Q]$; for example we will say that $\sigma[Q_i, Q_j]$ is some clique $Q$ to mean that the $Q$ is the clique whose identifier is $\sigma[id[Q_i], id[Q_j]]$ and we will write $n[u, Q]$ for $n[u, id[Q]]$. Moreover, we will say that some clique $Q$ belongs to a family ${\cal A}$ if the identifier of $Q$ belongs to ${\cal A}$.

\begin{Theorem}\label{DataStructureProperties}
The above data structures have the following properties:
\begin{description}
   \item[\emph{(i)}] for each $u \in V$, $|{\cal F}_u| \le 1$; 
   \item[\emph{(ii)}] for each node $u \in S$, the cliques in the family ${\cal C}_u$ belong to $\{C_u, \bar C_u\}$ and the family ${\cal F}_u$ is empty;
   \item[\emph{(iii)}] for each bound node $u \in V$ with $S(u) = \{s, t\}$, the cliques in the family ${\cal C}_u$ belong to $\{C_s, \bar C_s, C_t, \bar C_t\}$ and the family ${\cal F}_u$ is empty;
   \item[\emph{(iv)}] for each free node $u$ with $S(u) = \{s\}$, the cliques in the family ${\cal C}_u$ belong to $\{C_s, \bar C_s\}$;
   \item[\emph{(v)}] the families $\{{\cal C}_u : u \in V\}$, $\{{\cal F}_u : u \in V\}$, \{${\cal D}_s : s \in S\}$ and $\{{\cal D}_{st} : s, t \in S\}$ can be constructed in overall time ${\cal O}(|V|^2)$;
   \item[\emph{(vi)}]  for each $u \in V$, the family ${\cal B}_u$ contains at most $10$ sets;
   \item[\emph{(vii)}] the list ${\cal B}$ contains ${\cal O}(|V|)$ sets and can be constructed in ${\cal O}(|V|^2)$ time along with the matrix $\sigma[\cdot, \cdot]$ and the families ${\cal B}_u$ for each $u \in V$;
   \item[\emph{(viii)}] the matrix $n[\cdot, \cdot]$ can be computed in ${\cal O}(|V|^2)$ time.
\end{description}
\end{Theorem}

\noindent
\emph{Proof}. Properties~\emph{(i)}--\emph{(iv)} easily follow from the definition. Property~\emph{(v)} can be proved by the following procedure: first, initialize empty lists of identifiers; scan each clique $Q \in {\cal C} \subseteq {\cal B}$ and, for each $u \in Q$, add the identifier of $Q$ to ${\cal C}_u$; scan each clique $Q \in {\cal F} \subseteq {\cal B}$ and, for each $u \in Q$, add the identifier of $Q$ to ${\cal F}_u$; scan each clique $Q \in {\cal F} \subseteq {\cal B}$ and, if $Q$ is a trivial F-clique in $W(s, t)$, add the identifier of $Q$ to ${\cal D}_{st}$ otherwise, mark in ${\cal O}(|V|)$ time all the nodes in $S$ and, for each $u \in Q$ with the property that $S(u)$ is marked, add the identifier of $Q$ to ${\cal D}_{S(u)}$ and unmark $S(u)$. Since each node in $V$ belongs to at most four cliques in ${\cal C} \cup {\cal F}$ (properties~\emph{(i)}--\emph{(iv)}) the claim follows. To prove property~\emph{(vi)} we consider three cases. If $u \in S$ then, by property~\emph{(ii)}, $u$ is contained in at most two cliques of ${\cal C}$, in their intersection and in no clique of ${\cal F}$; hence $|{\cal B}_u| \le 3$. If $u$ is a bound node then, by property~\emph{(iii)}, $u$ is contained in at most four cliques in ${\cal C}$, in their intersections and in no clique of ${\cal F}$; in this case $|{\cal B}_u| \le 10$. Finally, if $u$ is a free node then, by properties~\emph{(i)} and \emph{(iv)}, $u$ is contained in at most two cliques of ${\cal C}$, one clique of ${\cal F}$ and their intersections. In addition $u$ is contained in the unique similarity class $F(S(u))$ and hence in at most one intersection of a clique in ${\cal F}$ with a free similarity class; in this case $|{\cal B}_u| \le 7$. Hence, for each $u \in V$, $|{\cal B}_u| \le 10$ as claimed. 

\medskip\noindent
To prove property~\emph{(vii)} we first show that ${\cal B}$, $\sigma[\cdot, \cdot]$ and ${\cal B}_u$ (for all $u \in V$) can be constructed in ${\cal O}(|V|^2)$ time. To this purpose, we first initialize ${\cal B} := {\cal C} \cup {\cal F} \cup \{F(s): s \in S\}$; by Theorem~\ref{SCoverFreeComponentComplexity} this can be done in ${\cal O}(|V|^2)$ time. Then, for each clique $Q \in {\cal C} \cup {\cal F} \cup \{F(s): s \in S\}$ we set $\sigma[Q, Q] := Q$ and, for each pair of distinct cliques $Q_i, Q_j \in {\cal C} \cup {\cal F} \cup \{F(s): s \in S\}$, we initialize a set $Q_{ij} := \emptyset$ and let $\sigma[Q_i, Q_j] := Q_{ij}$. Finally, for all $u \in V$, we initialize ${\cal B}_u$ as an empty list. The above initialization requires ${\cal O}(|V|^2)$ time. Subsequently, we scan the nodes $u \in V$ and do the following: for each pair of distinct cliques $Q_i, Q_j \in {\cal C}_u \cup {\cal F}_u$ we add $u$ to $Q_{ij} = \sigma[Q_i, Q_j]$ and add $Q_{ij}$ to the family ${\cal B}_u$. Moreover, 
if ${\cal F}_u$ is non-empty,
we let $Q_i$ be the unique (by property~\emph{(i)}) clique in ${\cal F}_u$, $Q_j = F(S(u))$ and again add $u$ to $Q_{ij} = \sigma[Q_i, Q_j]$ and $Q_{ij}$ to ${\cal B}_u$. Since by property~\emph{(vi)} there are at most $10$ cliques to be added to ${\cal B}_u$, the complete scan of $V$ can be performed in ${O}(|V|)$ time. Finally, ${\cal B}$ can be completed as follows: for each pair $Q_i, Q_j$ of distinct cliques in ${\cal C} \cup {\cal F} \cup \{F(s): s \in S\}$, if $Q_{ij} = \sigma[Q_i, Q_j]$ is non-empty add it to ${\cal B}$, otherwise set $\sigma[Q_i, Q_j] := nil$. This can be done in ${\cal O}(|V|^2)$ time. To prove that the list ${\cal B}$ contains ${\cal O}(|V|)$ non-empty sets it is sufficient to observe that it is composed by the cliques in $\cup_{u \in V} {\cal B}_u$ and that each ${\cal B}_u$ contains a constant number of cliques.

\medskip\noindent
To prove property~\emph{(viii)} we first let $n[u, Q] := 0$ for each node $u \in V$ and each clique $Q \in {\cal B}$. Then, for each edge $uv \in E$, each pair of cliques $Q_u \in {\cal B}_u$ and $Q_v \in {\cal B}_v$ (there are ${\cal O}(1)$ such cliques by property~\emph{(vi)}), increment both $n[u, Q_v]$ and $n[v, Q_u]$. Evidently the computation produces the desired matrix and can be carried out in overall time ${\cal O}(|V|^2)$.
\done

\medskip\noindent
\textsl{
\small
In figure~\ref{FigIrregular}, ${\cal B} = \{ \{1, 2\}, \{2, 3, 4\}, \{9, 10, 11\}, \{2\} \}$, $A = id[\{1, 2\}]$, $B = id[\{2, 3, 4\}]$, $C = id[\{9, 10, 11\}]$, $D = id[\{2\}]$, ${\cal B}_2 = \{A, B, D\}$, $\sigma[A, B] = D$, $\sigma[A, C] = nil$, $n[6, B] = 2$, $n[5, B] = 1$.
}

\begin{Theorem}\label{FindingPartitionCliqueComplexity}
The family ${\cal S}$ can be constructed in ${\cal O}(|V|^2)$ time.
\end{Theorem}

\noindent
\emph{Proof.} The overall procedure for constructing ${\cal S}$ can be described as follows: We first let $\bar{\cal S} := {\cal C} \cup {\cal F} \subseteq {\cal B}$, then remove from $\bar{\cal S}$ the cliques which are not weakly normal. Hence $\bar{\cal S}$ will contain the set of weakly normal cliques in ${\cal C} \cup {\cal F}$. Finally, we let ${\cal S} := \bar{\cal S}$ and, for each pair $Q_1, Q_2 \in \bar{\cal S}$ such that $N(Q_1 \cap Q_2) \not\subseteq Q_1 \cup Q_2$, we remove $Q_1$ and $Q_2$ from ${\cal S}$. We now show that the above operations can be carried out in overall ${\cal O}(|V|^2)$ time. 

\medskip\noindent
\emph{Claim (i). For any $s \in S$, if there exists a non-trivial F-clique $Q \in {\cal D}_s$ and two $Q$-close nodes $x, y$ with a common neighbor $z \in F(s) \cap Q$ then $x$ and $y$ belong to $N(s)$ and are $\bar Q$-close for every F-clique $\bar Q \in {\cal D}_s$.}

\smallskip\noindent
\emph{Proof.} Assume first that $x$ is complete to $Q \setminus F(s)$, let $\bar x \in Q \cap F(s)$ be a node not adjacent to $x$ (it exists since $Q$ is maximal) and let $t, q$ be distinct stable nodes in $N(Q \setminus F(s))$. Let $z_1 \in Q \cap F(t)$ and $z_2 \in Q \cap F(q)$ be nodes adjacent to $x$. It follows that $x$ is adjacent to $t$ (otherwise $(z_1: x, \bar x, t)$ would be a claw) and to $q$ (otherwise $(z_2: x, \bar x, q)$ would be a claw). But then $(x: t, q, z)$ is a claw in $G$, a contradiction. Hence, $x$ and analogously $y$ are not complete to $Q \setminus F(s)$. Consequently, there exist $\bar x, \bar y \in Q \setminus F(s)$ with $x \bar x \notin E$ and $y \bar y \notin E$. Moreover, by claw-freeness, we have $x \bar y, y \bar x \in E$. Finally, $xs \in E$ (otherwise $(z: x, \bar x, s)$ would be a claw) and, analogously, $ys \in E$. Let $\bar z$ be a node in $\bar Q \cap F(s)$. Since $S$ is canonical, $z \bar z \in E$. Moreover, since $\bar z \notin Q$, we have $\bar z \bar x, \bar z \bar y \notin E$, otherwise $\bar z$ would belong to the connected component of the free dissimilarity graph containing $\bar x$ and $\bar y$, a contradiction. In addition, $x$ does not belong to $\bar Q$ for, otherwise, it should be free and, being adjacent to $\bar y$, it would belong to $Q$, a contradiction. Analogously, $y \notin \bar Q$. Finally, we have $x \bar z \in E$ (otherwise $(z: x, \bar x, \bar z)$ would be a claw) and $y \bar z \in E$ (otherwise $(z: y, \bar y, \bar z)$ would be a claw). But then $x$ and $y$ are $\bar Q$-close and the claim follows.

\noindent
\emph{End of Claim (i).}

\medskip\noindent
\emph{Claim (ii). For any pair $s, t \in S$, a trivial F-clique $Q \in {\cal D}_{st}$ is not weakly normal if and only if there exists a bound node $x \in W(s, t)$ adjacent to some node $z \in Q \cap F(s)$ and to some node $v \in Q \cap F(t)$ and a node $y \in (N(s) \cup N(t)) \setminus Q$ adjacent either to $z$ or to $v$ and non-adjacent to $x$.}

\medskip
\begin{figure}[htbp]
\centering
\begin{tikzpicture}
\GraphInit[vstyle=Normal]

\tikzstyle{VertexStyle}=[shape = circle, fill = black, minimum size = 20pt, text = white, draw]
\Vertex[L=$t$, x=0, y=2]{t}
\Vertex[L=$s$, x=7, y=2]{s}

\tikzstyle{VertexStyle}=[shape = circle, fill = white, minimum size = 20pt, text = black, draw]
\Vertex[L=$x$, x=4, y=0.3]{x}
\Vertex[L=$v$, x=2.5, y=1.5]{v}
\Vertex[L=$z$, x=4.5, y=1.5]{z}
\Vertex[L=$u$, x=2.5, y=2.5]{u}
\Vertex[L=$q$, x=4.5, y=2.5]{q}
\Vertex[L=$y$, x=3, y=3.7]{y}

\tikzstyle{VertexStyle}=[shape = circle, minimum size = 17pt, draw]
\Vertex[L=$v$, x=2.5, y=1.5]{}
\Vertex[L=$z$, x=4.5, y=1.5]{}
\Vertex[L=$u$, x=2.5, y=2.5]{}
\Vertex[L=$q$, x=4.5, y=2.5]{}

\Edge(t)(u) \Edge(t)(x) \Edge(t)(v) \Edge(t)(y)
\Edge(s)(x) \Edge(s)(z) \Edge(s)(q) \Edge(s)(y)
\Edge(v)(u) \Edge(v)(q) \Edge(v)(z) \Edge(v)(x)
\Edge(u)(q) \Edge(u)(z) \Edge(u)(x)
\Edge(x)(z)
\Edge(y)(z) \Edge(y)(q)
\Edge(q)(z)

\end{tikzpicture}
\captionsetup{justification=centering, margin=2cm}
\caption{A trivial F-clique $\{u, v, q, z\}$ which is not weakly normal}
\label{FigTrivialFClique}
\end{figure}
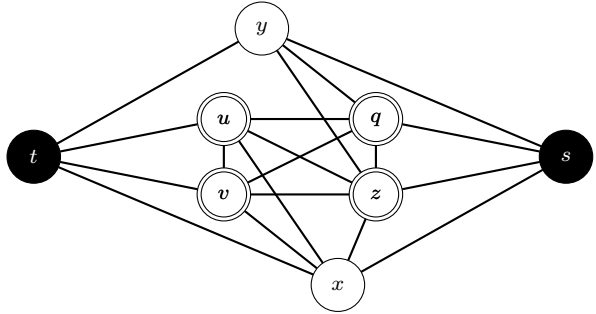

\smallskip\noindent
\emph{Proof.} If there exist a bound node $x \in W(s, t)$ and a node $y \in (N(s) \cup N(t)) \setminus Q$ both adjacent to a node $z \in Q$ with $xy \notin E$ then $Q$ is trivially not weakly normal.

\smallskip\noindent
On the other hand, if $Q$ is not weakly normal then there exist two nodes $x$ and $y$ not in $Q$ having a common neighbor $z$ in $Q$. Without loss of generality, assume $z \in Q \cap F(s)$. Moreover, if both $x$ and $y$ are non-adjacent to some node $v \in Q \cap F(t)$, then $(z: v, x, y)$ is a claw, a contradiction. Hence, without loss of generality, we can assume that $x$ is adjacent to some node $v \in Q \cap F(t)$. If $x$ is free then either $xz$ or $xv$ is an edge in the free dissimilarity graph and $x$ belongs to $Q$, a contradiction. It follows that $x$ is bound and, since it is adjacent to $z \in F(s)$ and $v \in F(t)$, by claw-freeness it must be adjacent to both $s$ and $t$ and so belongs to $W(s, t)$. Now, if $y$ is non-adjacent to $v$ then it must be adjacent to $s$ (otherwise $(z: s, y, v)$ would be a claw in $G$). On the other hand, if $y$ is adjacent to $v$ the same argument used for $x$ shows that $y$ is a bound node in $W(s, t)$. In both cases $y$ belongs to $(N(s) \cup N(t)) \setminus Q$ and the claim follows.

\noindent
\emph{End of Claim (ii).}

\medskip\noindent
\emph{Claim (iii). For any pair $s, t \in S$, if $|{\cal D}_{st}| \ge 2$ then a trivial F-clique $Q \in {\cal D}_{st}$ is not weakly normal if and only if there exists a bound node $x \in W(s, t)$ adjacent to both $Q \cap F(s)$ and $Q \cap F(t)$.}

\smallskip\noindent
\emph{Proof.} Assume first that $Q$ is not weakly normal. By Claim~\emph{(ii)} there exists a bound node $x \in W(s, t)$ adjacent to some node $z \in Q \cap F(s)$ and to some node $v \in Q \cap F(t)$, so the claim follows.

\smallskip\noindent
Suppose now that there exists a bound node $x \in W(s, t)$ adjacent to a node $z \in Q \cap F(s)$ and a node $v \in Q \cap F(t)$. Let $Q' \ne Q$ be another clique of ${\cal D}_{st}$ and let $y \in Q'$ be a node non-adjacent to $x$. Such a node exists because, otherwise, $Q'$ would not be maximal. Without loss of generality, assume $y \in Q' \cap F(s)$. Since $y$ and $z$ belong to the clique $F(s)$ ($S$ is canonical) we have that $z \in Q$ is a common neighbor of $x, y \in N(Q)$, hence $Q$ is not weakly normal and the claim follows.

\noindent
\emph{End of Claim (iii).}

\medskip\noindent
\emph{Claim (iv). If $Q$ belongs to $\bar{\cal S}$ and $u, v \in N(Q)$ are non-adjacent then $u$ and $v$ are $Q$-close if and only if $n[u, Q] + n[v, Q] > |Q|$.}

\smallskip\noindent
\emph{Proof.} By claw-freeness two non-adjacent nodes $u, v \in N(Q)$ with a common neighbor in $Q$ have the property that $N(\{u, v\}) \supseteq Q$ and hence $n[u, Q] + n[v, Q] > |Q|$. The converse is obvious.

\noindent
\emph{End of Claim (iv).}

\medskip\noindent
Now, we can remove from $\bar{\cal S}$ the F-cliques which are not weakly normal. First, for each trivial F-clique $Q$, letting $F(s)$ and $F(t)$ be the free similarity classes intersecting $Q$, we check whether $|{\cal D}_{st}| \ge 2$ and in this case remove $Q$ if there exists a bound node in $W(s, t)$ adjacent to both $Q_s = Q \cap F(s)$ and $Q_t = Q \cap F(t)$. Observe that the cliques $Q_s$ and $Q_t$ belong to ${\cal B}$ and can be retrieved as $Q_s := \sigma[Q, F(s)]$ and $Q_t := \sigma[Q, F(t)]$ so, for any bound node $x \in W(s, t)$, the check can be carried out by verifying whether $n[x, Q_s] \ne 0$ and $n[x, Q_t] \ne 0$. Hence, the above eliminations can be carried out in overall time ${\cal O}(|V|^2)$ and, by Claim~\emph{(iii)}, we have that the resulting $\bar{\cal S}$ still contains all the weakly normal cliques in ${\cal C} \cup {\cal F}$. Now each trivial F-clique $Q$ in $\bar{\cal S}$ which is not weakly normal is contained in some wing $W(s, t)$ satisfying $|{\cal D}_{st}| = 1$. By Claim~\emph{(ii)} $Q$ is characterized by the property that there exists a bound node $x \in W(s, t)$ and a non-adjacent node $y \in (N(s) \cup N(t)) \setminus Q$ with a common neighbor in $Q$. Hence, to remove such cliques we do the following. For each bound node $x$, let $W(s, t)$ be the wing containing it; we check whether $|{\cal D}_{st}| = 1$. If this is the case, we let $Q$ be the unique clique in ${\cal D}_{st}$ and mark all the nodes in $|V|$. Then, we unmark each node $v \in Q \cup N[x]$. 
Finally, for every node $y \in (N(s) \cup N(t))$, we check whether $y$ is marked (i.e.\ $y \in (N(s) \cup N(t)) \setminus (Q \cup N[x])$) and $n[x, Q] + n[y, Q] > |Q|$. By Claim~\emph{(iv)}, the non-adjacent nodes $x, y \in N(Q)$ have a common neighbor in $Q$ if and only if $n[x, Q] + n[y, Q] > |Q|$, so by such a procedure we can find (and remove from $\bar{\cal S}$) all the remaining trivial F-cliques which are not weakly normal. The above computations can be carried out in overall time ${\cal O}(|V|^2)$.

\medskip\noindent
Now, we can remove from $\bar{\cal S}$ the non-trivial F-cliques which are not weakly normal. 
Each one of them, say $Q$, contains some node $z$ which is the common neighbor of two non-adjacent nodes $x$ and $y$ in $N(Q)$. 
By Claim~\emph{(i)}, $x$, $y$ and $z$ are adjacent to a common stable node $s \in S$. Hence, to remove all the non-trivial F-cliques which are not weakly normal we do the following. For each $s \in S$ we select one F-clique $Q \in {\cal D}_s$ (if any) and, 
for each pair $x, y$ of non-adjacent nodes in $N(s) \setminus Q$, 
we check whether $n[x, Q] + n[y, Q] > |Q|$. If such a pair exists, by Claim~\emph{(i)}, we have that each F-clique in ${\cal D}_s$ is not weakly normal and can be removed. Otherwise, no F-clique in ${\cal D}_s$ contains, in $F(s)$, the common neighbor of two non-adjacent nodes. Since, by claw-freeness, each pair $x, y \in V$ is adjacent to at most two nodes in $S$, we have that the above computations can be carried out in overall time ${\cal O}(|V|^2)$. Moreover, the above procedure removes all the non-trivial F-cliques which are not weakly normal.

\medskip\noindent
Now, to remove the cliques in $\bar{\cal S} \cap {\cal C}$ which are not weakly normal we check, for each pair of non-adjacent nodes $x, y \in V$, whether they have a common neighbor in some clique $Q \in \bar{\cal S} \cap {\cal C}$ not containing $x$ and $y$ and, in such a case, remove $Q$ from $\bar{\cal S}$. To assess the complexity of this operation suppose first that both $x$ and $y$ are bound and belong, respectively, to the possibly coincident wings $W(s, t)$ and $W(u, v)$. Assume that there exists a clique $Q \in \bar{\cal S} \cap {\cal C}$ containing a common neighbor $q$ of $x$ and $y$. Let $z$ be the node in $Q \cap S$ and observe that $z$ is in $\{s, t, u, v\}$, otherwise $(q: x, y, z)$ would be a claw in $G$. It follows that $Q$ belongs to one of the pairs $(C_s, \bar C_s)$, $(C_t, \bar C_t)$, $(C_u, \bar C_u)$, $(C_v, \bar C_v)$ and hence for each pair $x, y$ of non-adjacent bound nodes we have to test at most eight cliques. Similar arguments show that if $x$ and/or $y$ is free 
or bound 
then we have to test less than eight cliques. Moreover, by  Claim~\emph{(iv)}, we can verify that the non-adjacent nodes $x, y$ have a common neighbor in $Q$ by checking whether $n[x, Q] + n[y, Q] > |Q|$. This implies that the overall check can be performed in ${\cal O}(|V|^2)$ time.

\medskip\noindent
Now, to conclude the construction of ${\cal S}$, we first mark with two labels (blue and red) the cliques in ${\cal B}$ belonging to $\bar{\cal S}$. This can be done in ${\cal O}(|V|)$ time. Then, for each node $u \in V$ and each clique $B \in {\cal B}$ with $B = Q_1 \cap Q_2$ and $Q_1, Q_2$ marked in red (i.e.\ $Q_1, Q_2 \in \bar{\cal S}$), we remove the blue label from both $Q_1$ and $Q_2$ if $n[u, B] \ge 1$ and $Q_1, Q_2 \notin {\cal B}_u$. Since, by Theorem~\ref{DataStructureProperties}, ${\cal B}$ contains ${\cal O}(|V|)$ elements and each family ${\cal B}_u$ ${\cal O}(1)$ elements, the overall check can be performed in ${\cal O}(|V|^2)$ time. Finally, in ${\cal O}(|V|)$ time, we construct ${\cal S}$ by adding all the cliques in $\{\cal B\}$ marked in blue. The theorem follows.
\done

\section{Ungluing of $S$-articulation cliques in ${\cal O}(|V|^2)$}

In this section we show that all the edges satisfying the conditions of Definition~\ref{Ungluing} can be removed in ${\cal O}(|V|^2)$ time. To this purpose we show that, for all cliques $Q \in {\cal S}$, we can determine the connected components of the rigid structure $G_R[Q]$ and, according to Definition~\ref{Ungluing}, remove any edge $uv$ such that $u$ and $v$ belong to different components of $G_R[Q]$ in overall time ${\cal O}(|V|^2)$.

\begin{Theorem}\label{FindingSoftCliquesComplexity}
The ungluing $G_{\cal S}$ of $G$ with respect to ${\cal S}$ can be constructed in ${\cal O}(|V|^2)$ time.
\end{Theorem}

\noindent
\emph{Proof.} By Theorem~\ref{FindingPartitionCliqueComplexity} the family ${\cal S}$ of $S$-articulation cliques can be constructed in ${\cal O}(|V|^2)$ time and contains ${\cal O}(|V|)$ elements. Then, in overall time ${\cal O}(|V|^2)$, we can construct the list $\{{\cal S}_u : u \in V\}$, where ${\cal S}_u$ is the family of cliques in ${\cal S}$ containing the node $u$. Observe that, by Theorem~\ref{NoThreeSArticulation}, $|{\cal S}_u| \le 2$. 

\medskip\noindent
We now compute the connected components of the rigid structure of the $S$-articulation cliques. To this purpose, for each node $u \in V$ and each clique $Q \in {\cal S}$ which does not contain $u$, let $Root[u, Q]$ be a node in $Q \cap N(u)$ which does not belong to any clique $Q' \in {\cal S} \setminus \{Q\}$ (if any). Moreover, for each clique $Q \in {\cal S}$, let $G_Q(Q, E_Q)$ be the spanning subgraph of $G[Q]$ with $xy \in E_Q$ if and only if either both $x$ and $y$ belong to $Q \cap Q_i$ for some $Q_i \in {\cal S} \setminus \{Q\}$ or $Q$ is the unique clique in ${\cal S}$ containing both $x$ and $y$ and there exists a node $u$ satisfying $x = Root[u, Q]$ and $y \in N(u)$. Observe that each edge $xy \in E_Q$ is rigid. In fact, if $x$ and $y$ belong to $Q \cap Q_i$ for some $Q_i \in {\cal S} \setminus \{Q\}$ then, by Theorem~\ref{NoThreeSArticulation}, $x$ and $y$ are not distinguished by any clique in ${\cal S}$ and, since $Q \ne Q_i$, $N(x) \cap N(y)$ is not a clique in $G$. If, on the other hand, $Q$ is the unique clique in ${\cal S}$ containing both $x$ and $y$ then $x$ and $y$ are not distinguished by any clique in ${\cal S}$. Moreover, there exists some node $u \notin Q$ satisfying $x = Root[u, Q] \in N(u)$ and $y \in N(u)$; hence $N(x) \cap N(y)$ is not a clique in $G$. In both cases $xy$ is a rigid edge. We have the following:

\medskip\noindent
\emph{Claim (i). The matrix $Root[\cdot, \cdot]$ and the graphs $G_Q$ ($Q \in {\cal S}$) can be computed in ${\cal O}(|V|^2)$ time.}

\smallskip\noindent
\emph{Proof.} For each $Q \in {\cal S}$ we initialize the graph $G_Q$ by letting $E_Q := \emptyset$ and, for each node $u \in V$, we let $Root[u, Q] := nil$. Now, for each edge $uv \in E$ we do the following. If ${\cal S}_v$ contains a single clique $Q^v$ not in ${\cal S}_u$ then if $Root[u, Q^v] = nil$, we let $Root[u, Q^v] := v$, otherwise we add to $E_Q$ the edge $(v, Root[u, Q^v])$; note that, in this case, $Root[u, Q^v] = v'$ for some $uv' \ne uv$ and hence $v$ is different from $Root[u, Q^v]$. Analogously, if ${\cal S}_u$ contains a single clique $Q^u$ not in ${\cal S}_v$ then we procede as above by interchanging the roles of $u$ and $v$. Finally, if both ${\cal S}_u$ and ${\cal S}_v$ contain the same pair of cliques $Q_1$ and $Q_2$ then we add the edge $uv$ both to $E_{Q_1}$ and to $E_{Q_2}$. Evidently the above procedure produces the matrix $Root[\cdot, \cdot]$ and the graphs $G_Q$ for $Q \in {\cal S}$ in overall time ${\cal O}(|V|^2)$.

\noindent
\emph{End of Claim (i).}

\medskip\noindent
\emph{Claim (ii). For each $Q \in {\cal S}$ the connected components of $G_R[Q]$ coincide with the connected components of $G_Q$.}

\smallskip\noindent
\emph{Proof.} Since all the edges in $E_Q$ are rigid, we have that any connected component of $G_Q$ is contained in some connected component of $G_R[Q]$. We have to show that also the reverse is true. Hence suppose, by contradiction, that there exists a rigid edge $xy \in E$ with $x$ and $y$ in different connected components of $G_Q$. Observe that, for any clique $Q_i \in {\cal S} \setminus \{Q\}$, an edge $uv$ with $u \in Q \cap Q_i$ and $v \in Q \setminus Q_i$ is not rigid ($u$ and $v$ are distinguished by $Q_i$) and an edge $uv$ with both $u$ and $v$ in $Q \cap Q_i$ belongs to $E_Q$, so we must have that $Q$ is the unique clique in ${\cal S}$ containing both $x$ and $y$. Let $u$ be a node in $N(Q)$ adjacent to both $x$ and $y$ (it exists since $xy$ is rigid). Let $z = Root[u, Q]$. If $z \equiv x$ or $z \equiv y$ then, by construction, $xy$ belongs to $E_Q$, a contradiction. It follows that $z \ne x, y$ and, by construction, the edges $xz$ and $yz$ belong to $E_Q$. But then $x$ and $y$ belong to the same connected component of $G_Q$, contradicting the assumption.

\noindent
\emph{End of Claim (ii).}

\medskip\noindent
Observe that the overall complexity of constructing $\bar G_Q$ for each $Q \in {\cal S}$ is ${\cal O}(|V|^2)$ (recall that any edge belongs to at most two $S$-articulation cliques). Hence the ungluing $G_{\cal S}$ of $G$ along with the corresponding partitions of the cliques $Q \in {\cal S}$ can be produced in ${\cal O}(|V|^2)$ time  and the theorem follows.
\done

\section{Conclusion}

\noindent
The results of the last section show that a generalization of the ungluing operation defined by Faenza, Oriolo and Stauffer \cite{FaenzaOS14} applies directly to the family of $S$-articulation cliques in a claw-free graph $G(V, E)$ and produces a collection of \{claw, net\}-free strips and strips with stability number at most three. By Theorems~\ref{FindingPartitionCliqueComplexity} and \ref{FindingSoftCliquesComplexity} the overall complexity of finding the $S$-articulation cliques and ungluing the graph $G$ is ${\cal O}(|V|^2)$. Moreover, the results of \cite{NobiliSassano16a} and \cite{NobiliSassano17a} allow us to solve the MWSS problem in ${\cal O}(|V(C)| \sqrt{|E(C)|})$ time in each \{claw, net\}-free strip $C$ with $\alpha(C) \ge 4$ and in ${\cal O}(|E(C)| \log |V(C)|)$ time in each strip $C$ with $\alpha(C) \le 3$. It follows that the overall time complexity of solving the MWSS problem in all the strips is ${\cal O}(|V|^2 \log |V|)$. In \cite{FaenzaOS14} (Theorem~2.10) Faenza, Oriolo and Stauffer have shown that the MWSS problem on a graph $G(V, E)$ that is the composition of some set of strips can be solved in ${\cal O}(|V|^2 \log |V|)$ plus the time of solving the same problem in all the strips. Consequently, the results of our three papers show that the MWSS problem on claw-free graphs can be solved in ${\cal O}(|V|^2 \log |V|)$ time. This improves with respect to the ${\cal O}(|V| (|V| \log |V| + |E|))$ bound achieved in \cite{FaenzaOS14} and is aligned to the complexity of the best known algorithm which solves the MWSS problem in line graphs \cite{Gabow90}. As Manfred Padberg conjectured in 1983, \emph{solving the maximum weight stable set problem on claw-free graphs is not harder than solving the matching problem}.


\begin{thebibliography}{10}
\providecommand{\url}[1]{{#1}}
\providecommand{\urlprefix}{URL }
\expandafter\ifx\csname urlstyle\endcsname\relax
  \providecommand{\doi}[1]{DOI~\discretionary{}{}{}#1}\else
  \providecommand{\doi}{DOI~\discretionary{}{}{}\begingroup
  \urlstyle{rm}\Url}\fi

\bibitem{BD83}
Ball, M.O., Derigs, U.: An analysis of alternative strategies for implementing
  matching algorithms.
\newblock Networks \textbf{13}(4), 517--549 (1983)

\bibitem{ChudSeym}
Chudnovsky, M., Seymour, P.D.: The structure of claw-free graphs.
\newblock In: Surveys in Combinatorics, pp. 153--171 (2005)

\bibitem{FaenzaOS14}
Faenza, Y., Oriolo, G., Stauffer, G.: Solving the weighted stable set problem
  in claw-free graphs via decomposition.
\newblock J. {ACM} \textbf{61}(4), 20 (2014)

\bibitem{Fouquet}
Fouquet, J.: {A strengthening of Ben Rebea's lemma.}
\newblock J. Comb. Theory, Ser. B \textbf{59}(1), 35--40 (1993)

\bibitem{Gabow90}
Gabow, H.N.: Data structures for weighted matching and nearest common ancestors
  with linking.
\newblock In: SODA, pp. 434--443 (1990)

\bibitem{GMG82}
Galil, Z., Micali, S., Gabow, H.N.: Priority queues with variable priority and
  an {O(EV} log {V)} algorithm for finding a maximal weighted matching in
  general graphs.
\newblock In: 23rd Annual Symposium on Foundations of Computer Science,
  Chicago, Illinois, USA, 3-5 November 1982, pp. 255--261 (1982)

\bibitem{KKM00}
Kloks, T., Kratsch, D., M{\"u}ller, H.: Finding and counting small induced
  subgraphs efficiently.
\newblock Inf. Process. Lett. \textbf{74}(3-4), 115--121 (2000)

\bibitem{LovaszPlummer}
Lov\'asz, L., Plummer, M.: {Matching theory.}
\newblock {Annals of Discrete Mathematics, 29. North-Holland Mathematics
  Studies, 121. Amsterdam etc.: North-Holland. XXXIII, 544 p.} (1986)

\bibitem{Minty}
Minty, G.J.: {On maximal independent sets of vertices in claw-free graphs.}
\newblock J. Comb. Theory, Ser. B \textbf{28}, 284--304 (1980)

\bibitem{NobiliSassano16a}
Nobili, P., Sassano, A.: An ${O}(n \sqrt{m})$ algorithm for the weighted stable
  set problem in \{Claw, Net\}-free graphs with $\alpha({G}) \geq 4$.
\newblock Discret. Optim. \textbf{19}(C), 63--78 (2016).
\newblock \doi{10.1016/j.disopt.2016.01.002}.
\newblock \urlprefix\url{http://dx.doi.org/10.1016/j.disopt.2016.01.002}

\bibitem{NobiliSassano17a}
Nobili, P., Sassano, A.: An ${O}(m \log n)$ algorithm for the weighted stable
  set problem in claw-free graphs with $\alpha({G}) \le 3$.
\newblock Mathematical Programming \textbf{164}(1), 157--165 (2017).
\newblock \doi{10.1007/s10107-016-1080-9}.
\newblock \urlprefix\url{https://doi.org/10.1007/s10107-016-1080-9}

\bibitem{Schrijver}
Schrijver, A.: Combinatorial Optimization: Polyhedra and Efficiency.
\newblock Springer-Verlag, Berlin Hiedelberg (2003)

\end{thebibliography}
\end{document}